\documentclass{vldb}
\pdfoutput=1
\usepackage{graphicx}
\usepackage{algorithmic}
\usepackage{algorithm}
\usepackage{multirow}
\usepackage{balance}  
\usepackage{ifpdf,theorem,tabularx,array,booktabs}
\usepackage{balance}
\usepackage{fixltx2e}

\newcolumntype{C}{>{\centering\arraybackslash} X }

\newcounter{enum}

\newcounter{enum2}

\newenvironment{packed_item}{
\begin{list}{$\bullet$}{
  \setlength{\topsep}{0pt}
  \setlength{\itemsep}{-1.5pt}
  \setlength{\parskip}{1pt}
  \setlength{\labelwidth}{15pt}
  \setlength{\leftmargin}{10pt}
  \setlength{\itemindent}{0pt}}
}{\end{list}}

\newcounter{lett}

\newcounter{rom}

\theoremstyle{plain}
\theoremheaderfont{\scshape}

\theorembodyfont{\normalfont}

\newcommand{\hide}[1]{}
\newcommand{\TR}[1]{}

\usepackage{fancyhdr}
\usepackage[mmddyy]{datetime}

\begin{document}


\title{Getting It All from the Crowd}



%
%
%
%

\numberofauthors{2}
\author{
\begin{tabular}{cccc}\\
Beth Trushkowsky & Tim Kraska  & Michael J. Franklin & Purnamrita Sarkar
\end{tabular}
\\
\begin{tabular}{c}\\
\affaddr{AMPLab, UC Berkeley} \\
\affaddr{\{trush, kraska, franklin, psarkar\}@cs.berkeley.edu}
\end{tabular}
}


%
%


\maketitle

\begin{abstract}
Hybrid human/computer systems promise to greatly expand the usefulness of query processing by incorporating the crowd for data gathering and other tasks.   Such systems raise many database system implementation questions.  Perhaps most fundamental is that the closed world assumption underlying relational query semantics does not hold in such systems.  As a consequence the meaning of even simple queries can be called into question. Furthermore query progress monitoring becomes difficult due to non-uniformities in the arrival of crowdsourced data and peculiarities of how people work in crowdsourcing systems.   To address these issues, we develop statistical tools that enable users and systems developers to reason about tradeoffs between time/cost and completeness.  These tools can also help drive query execution and crowdsourcing strategies.  We evaluate our techniques using experiments on a popular crowdsourcing platform.
\end{abstract}

\section{Introduction}
Advances in machine learning, natural language processing, image understanding, etc.\ continue to expand the range of problems that can be addressed by computers.  But despite these advances, people still outperform state-of-the-art algorithms for many data-intensive tasks.  Such tasks typically involve ambiguity, deep understanding of language or context, or subjective reasoning.  

Crowdsourcing has emerged as a paradigm for leveraging human intelligence and activity at large scale.  Popular crowdsourcing platforms such as Amazon Mechanical Turk (AMT) provide access to hundreds of thousands of human workers via programmatic interfaces (APIs).  These APIs provide an intriguing new opportunity, namely, to create hybrid human/computer systems for data-intensive applications.  Such systems, could, to quote J.C.R. Licklider's famous 1960 prediction for man-computer symbiosis, ``...process data in a way not approached by the information-handling machines we know today.''~\cite{licklider}.

\subsection{Query Processing with Crowds}
\label{intro:hybridDBs}
Recently, a number of projects have begun to explore the potential of hybrid human/computer systems for database query processing.   These include CrowdDB~\cite{crowddb}, Qurk~\cite{qurk}, and sCOOP~\cite{scoop}.  In these systems, human workers can perform query operations such as subjective comparisons, fuzzy matching for predicates and joins, entity resolution, etc.  

For example, CrowdDB incorporates several SQL language extensions to involve people in query processing.   Of particular relevance to the work we present here, the CrowdDB Data Definition Language (DDL) includes the special keyword \texttt{``CROWD''} to indicate when missing values of existing records or entire missing rows of certain tables can be obtained via human input, say by posing jobs on a crowdsourcing platform such as AMT.  As shown in~\cite{crowddb}, these simple extensions can greatly extend the usefulness of a query processing system.

In an operator-based relational query engine, crowd processing can be encapsulated into operators that can be used along with traditional computer-based operators in query plans.  Of course, many challenges arise when adding people to query processing, due to the peculiarities in latency, cost, quality and predictability of human workers.  Such challenges impact nearly all aspects of database system design and implementation.  Data cleaning is also an issue.  Data obtained from the crowd must be validated, spelling mistakes must be fixed, duplicates must be removed etc.  Similar issues arise in data ingest for traditional database systems through ETL (Extract, Transform and Load) and data integration but techniques have also been developed specifically for crowdsourced input~\cite{panos_quality, automan, QoE, crowdtutorial}.  

The above concerns, while both interesting and important are not the focus of this paper.  Rather, we believe that there are more fundamental issues at play in such hybrid systems.  Specifically, when the crowd can augment the data in the database to help answer a query (as is enabled by CrowdDB's  \texttt{``CROWD''} keyword), the traditional \emph{closed-world assumption} on which relational database query processing is based, no longer holds.   This fundamental change calls into question the basic meaning of queries and query results in a hybrid human/computer database system.

\subsection{Can You Really Get it All?}
\label{intro:GettingAll}
In this paper, we consider a basic RDBMS operation, namely, enumerating the tuples in a relation.  Consider for example, 
a SQL query to count the records in a table \texttt{SELECT COUNT(*) FROM TABLE} (where the table has a primary key).   In a traditional RDBMS there is a single correct answer for this query, and it can be obtained by scanning the table, incrementing a counter for each record found, and  returning the count once all the records of the table have been read.  This approach works even for relations that are in reality unbounded, because the closed world assumption dictates that any records not present in the database at query execution time do not exist.  Of course, such limitations can be a source of frustration for users trying to obtain useful real-world information from database systems.

In contrast, in a crowdsourced system like CrowdDB, once the records in the stored table are exhausted, jobs can be sent to the crowd asking for additional records.  The question then arises as to when the query has been completed. Crowdsourced queries can be inherently ambiguous or effectively unbounded.  For example, consider a query to find a list of graduating Ph.D. students currently on the job market, or companies in California interested in green technology.  The queries do not have a known result cardinality, or even a unique, correct answer.  Thus, the meaning of even a simple enumeration query such as the \texttt{SELECT} query above becomes unclear.

Of course, in some cases, the cardinality of the relation being queried can be known or estimated \emph{a priori}, for example, a query asking for the names of the 50 US states.  Even for such queries, however, it is difficult to assess progress in terms of remaining time or cost, because answers arrive from the crowd in a non-uniform way.

To understand these issues, in this paper we address two fundamental questions: First, ``Is it really possible to `get it all from the crowd'?''  As the discussion above indicates, the answer to this question is: ``sometimes''. Thus, the second question we address is ``How should users think about enumeration queries in the open world of a crowdsourced database system?''.   For this second question, we develop statistical tools that enable users to reason about tradeoffs between time/cost and completeness and that can be used to drive query execution and crowdsourcing strategies.

\subsection{Counting Species}
\label{intro:SpeciesEstimation}
Consider the execution of a ``\texttt{SELECT *}'' query in a crowdsourced database system where workers are asked to provide individual records of the table.   For example, one could query for the names of the 50 US states using a microtask crowdsourcing platform like AMT by generating HITs (i.e., Human Intelligence Tasks) that would have workers provide the name of one or more states.  As workers return results, the system collects the answers, keeping a list of the unique answers (suitably cleansed) as they arrive. 
  
\begin{figure}[t]
	\centering
    \includegraphics[width=2.3in]{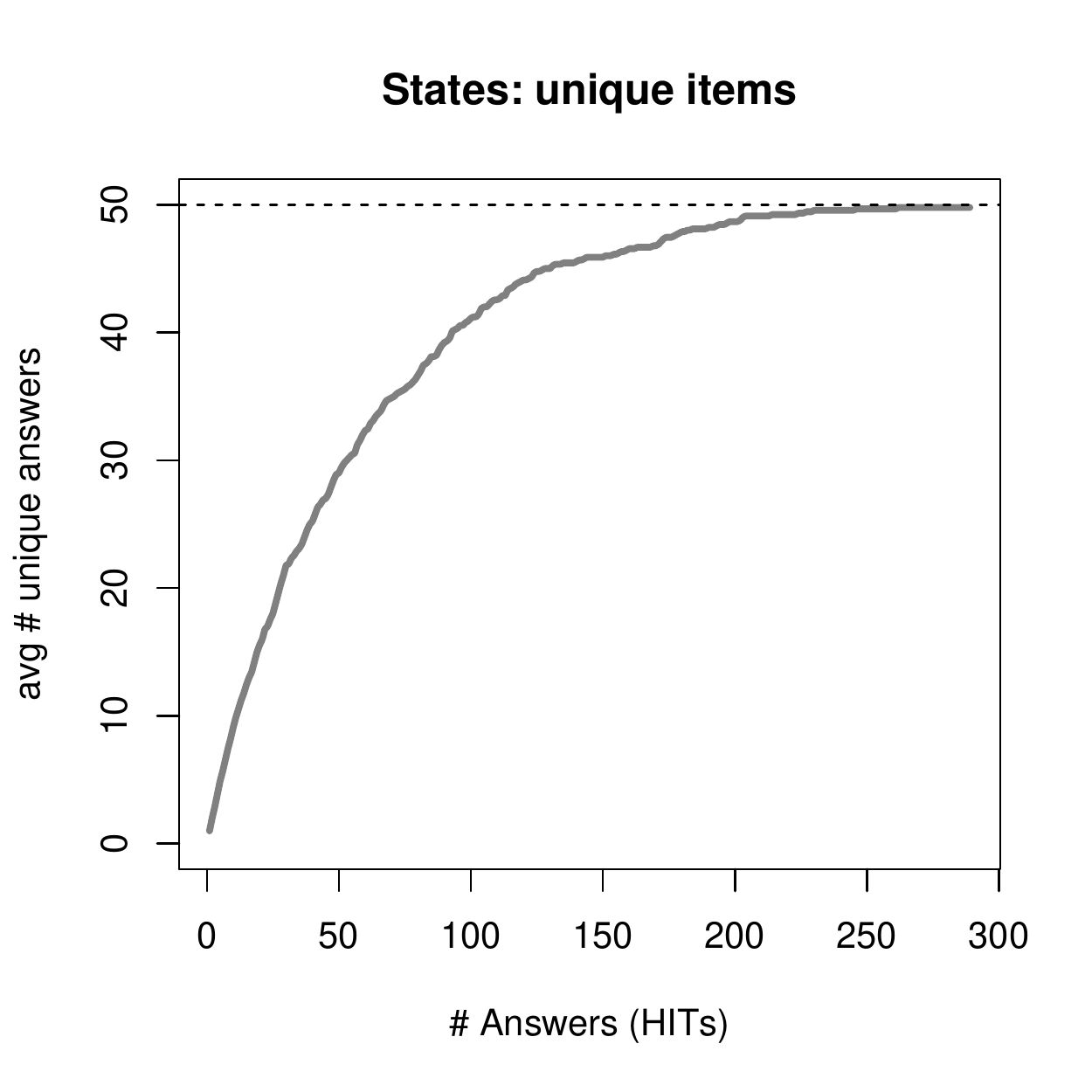}
\vspace*{-10pt}
	\caption{States experiments: number of unique items vs.\ total number of answers}
     \label{fig:states_sac}
\end{figure}

Figure~\ref{fig:states_sac} shows the results of running that query, with the number of unique answers received shown on the vertical axis, and the total number of answers received on the x-axis.  As would be expected, initially there is a high rate of arrival for previously unseen answers, but as the query progresses (and more answers have been seen) the arrival rate of new answers begins to taper off, until the full population (i.e., the 50 states, in this case) has been identified.

This behavior is well-known in fields such as biology and statistics, where this type of figure is known as the \emph{Species Accumulation Curve} (SAC)~\cite{colwell_review94}.  Imagine you were trying to count the number of unique species of animals on an island by putting out traps overnight, identifying the unique species found in the traps the next morning, releasing the animals and repeating this daily.   By observing the rate at which new species are identified over time, you can begin to infer how close to the true estimate of the number of species you are.   We can use similar reasoning to help understand the execution of set enumeration queries in a crowdsourced query processor.
\vspace{0.45in}

\subsection{Overview of the Paper}
\label{intro:Overview}
In this paper, we investigate the use of species (or ``classes'') estimation techniques from the statistics and biology literature for understanding and managing the execution of set enumeration queries in crowdsourced database systems.  We find that while the classical theory provides the key to understanding the meaning of such queries, there are certain peculiarities in the behavior of microtask crowdsourcing workers that require us to develop new methods to improve the accuracy of cardinality estimation and the quality of crowdsourced answers in this environment.  Furthermore, given the inherent ambiguity and unboundedness of many of the queries in a hybrid human/computer query processing system, we develop methods to leverage these techniques to help users make intelligent tradeoffs between time/cost and completeness.

To summarize, we make the following contributions:

\begin{packed_item}
		\item We apply species estimation algorithms in the new context of crowd-provided tuples to estimate result cardinality and query progress. 
		\item We develop new heuristics to improve these estimations in the presence of crowd-specific behaviors; namely, over-ambitious workers and workers using the same sequence of answers.
		\item We devise pay-as-you-go approaches to allow informed decisions about the cost/completeness tradeoff.
		\item We examine the effectiveness of our techniques via experiments using Amazon Mechanical Turk
	\end{packed_item}
	
\noindent The paper is organized as follows: 
In Section~\ref{sec:background} we describe the CrowdDB system and the use of species estimation in traditional closed-world database systems. 
Section~\ref{sec:cardest} evaluates different species estimation techniques in the context of crowdsourced queries. 
In Section~\ref{sec:sample_repl} we develop techniques to ameliorate the effect of over-ambitious workers. 
Section~\ref{sec:payg} introduces pay-as-you-go techniques. 
In Section~\ref{sec:listwalking} we present a new heuristic to detect the effects of workers using the same sequence of answers. 
Section~\ref{sec:related} covers related work and Section~\ref{sec:future} presents our conclusions and future work.


\vspace{0.5in}
\section{Background}
\label{sec:background}
In this section we describe the CrowdDB system, which serves as the context for this work. We then discuss related work on cardinality estimation in both the Statistics and Database Query Processing domains. 

\subsection{CrowdDB Overview}
\label{sec:background:crowddb}
CrowdDB is a hybrid human-machine database system that uses human input to process queries. 
CrowdDB currently supports two crowdsourcing platforms: AMT and our own mobile platform~\cite{CrowdDBVLDB11}. 
We focus on AMT in this paper, the leading platform for so-called microtasks.
Microtasks, also called Human Intelligence Tasks (HITs) in AMT, usually do not require any special training and do not take more than a few minutes to complete.
AMT provides a marketplace for microtasks that allows requesters to post HITs and workers to search for and work on HITs for a small reward, typically a few cents each.

\begin{figure}[bt]
	\centering
	\includegraphics[trim=0.2cm 0.2cm 0.2cm 0.2cm, width=7cm]{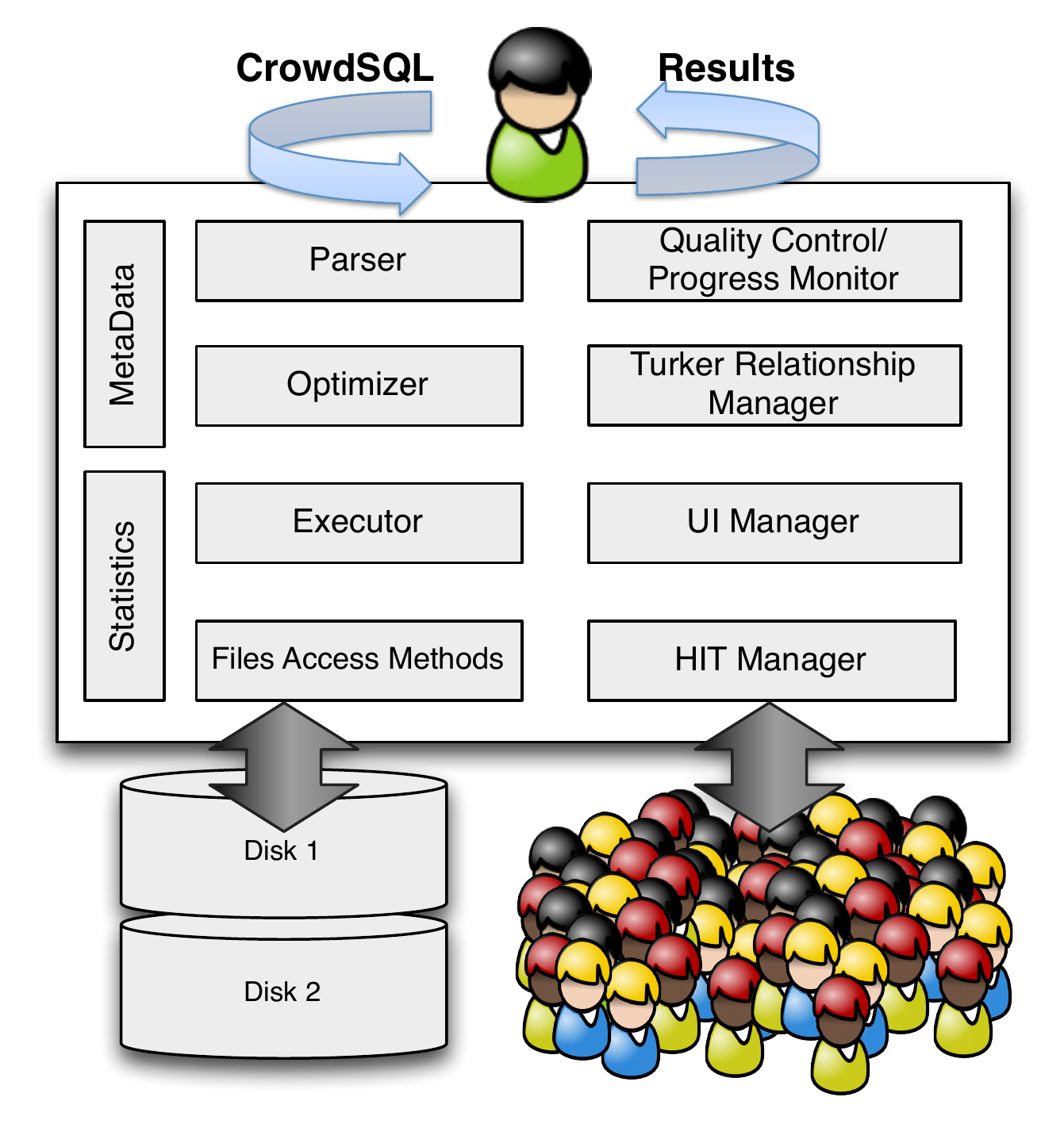}
	\vspace{-5pt}
	\caption{CrowdDB Architecture}
	\label{fig:arch}
	\vspace{-5pt}
\end{figure}

\begin{figure}[t]

	\centering
		\fbox{ 
    		\includegraphics[width=3in]{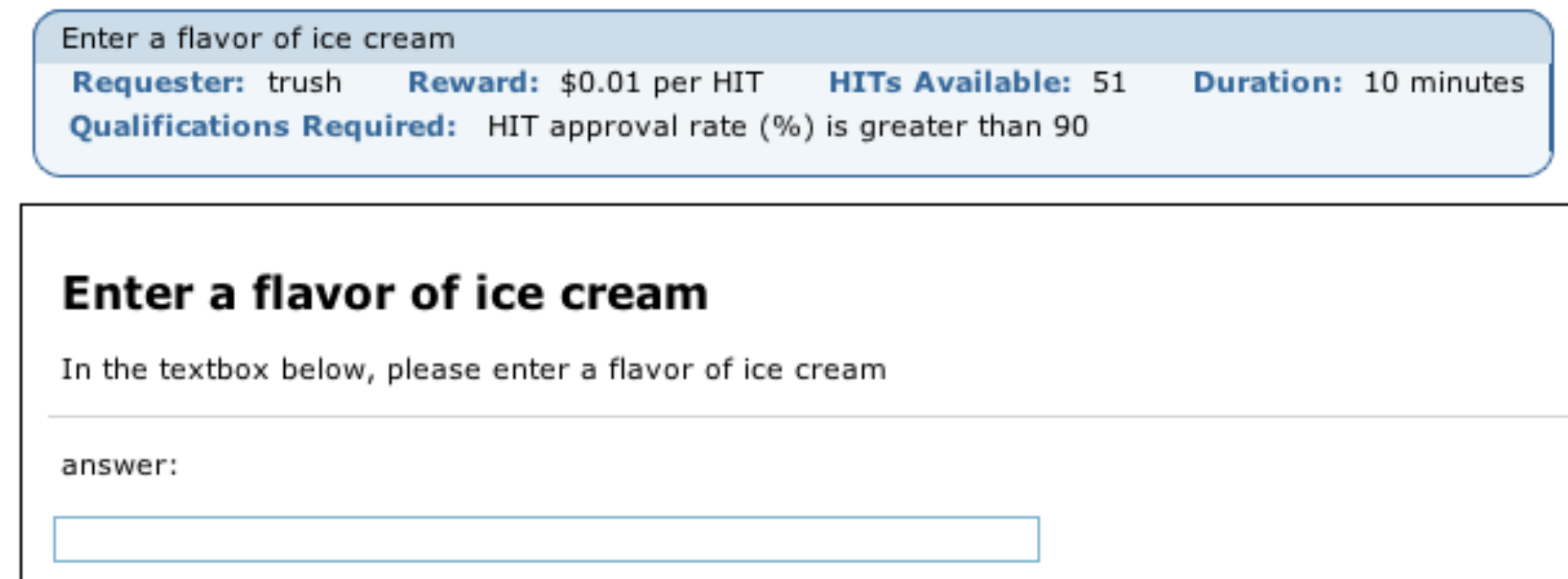}
		}
		\vspace{-5pt}
	\caption{Ice cream flavors task UI on AMT}
     \label{fig:ui}
	\vspace{-10pt}
\end{figure}

Figure~\ref{fig:arch} shows the architecture of CrowdDB.
CrowdDB incorporates traditional query compilation, optimization and execution components, which are extended to cope with human-generated input.  
In addition the system is extended with crowd-specific components, such as a user interface (UI) manager and quality control/progress monitor.
Users issue queries using CrowdSQL, an extension of standard SQL. 
CrowdDB automatically generates UIs as HTML forms based on the \texttt{CROWD} annotations and optional free-text annotations of columns and tables in the schema. 
Figure~\ref{fig:ui} shows an example HTML-based UI that would be presented to a worker for the following crowd table definition:
\begin{verbatim}
CREATE CROWD TABLE ice_cream_flavor {
     name VARCHAR PRIMARY KEY 
}
\end{verbatim}
Although CrowdDB supports alternate user interfaces (e.g., showing previously received answers), this paper focuses on a pure form of the ``getting it all'' question.  The use of alternative UIs is the subject of future work. 

During query processing, the system automatically posts one or more HITs using the AMT web service API and collects the answers as they arrive. 
After receiving the answers, CrowdDB performs simple quality control using quorum votes before it passes the answers to the query execution engine.
Finally, the system continuously updates the query result and estimates the quality of the current result based on the new answers. 
The user may thus stop the query as soon as the quality is sufficient or intervene if a problem is detected.
More details about the CrowdDB components and query execution are given in \cite{crowddb}.
We describe in this paper how the system can estimate completeness of the query result using algorithms from the species estimation literature.

\subsection{Cardinality Estimation}
\label{sec:bg_est}
To estimate progress as answers are arriving, the system needs an estimate of the result set's cardinality. 
We can tackle cardinality estimation by applying in a new context algorithms developed for estimating the number of species. 
In the species estimation problem \cite{bunge_review93,chao_review05}, an estimate of the number of distinct species is determined using observations of species in the locale of interest. 
These observations represent samples drawn from a probability distribution describing the likelihood of seeing each item. 
By drawing a parallel between observed species and answers received from the crowd, we can apply these techniques to reason about the result set size of a crowdsourced query. 


Work on distinct value estimation in traditional database systems has also looked into species estimation techniques to inform query optimization of large tables; tuples are sampled to estimate the number of distinct values present. 
Techniques used and developed in that literature leverage knowledge of the full table size, which is possible only because of the closed-world assumption. 
In the species estimation literature, the difference between these two scenarios is referred to as finite vs. infinite populations, which correspond to closed vs. open world, respectively. 

In \cite{haas}, Haas et. al. survey different estimators, several of which we also investigate in this paper.  
They do not use the algorithm we find superior because they observe it produced overly large estimates when used in the context of a finite population. 
Instead they propose a hybrid approach, choosing between the Shlosser estimator \cite{shlosser} and a version of the Jackknife estimator \cite{bo78} they modified to suit a finite population. 
The Jackknife technique is used for tables in which distinct values are uniformly distributed. 

This work was extended in \cite{charikar}, in which Charikar et. al. propose a different hybrid approach. 
They note a lack of analytic guarantees on errors in previous work, and derive a lower bound on error that an estimator should achieve. 
They then show that their algorithm is superior to Shlosser in the non-uniform case, substituting it in the hybrid approach from \cite{haas}. 
Unfortunately, both the error bounds and developed estimators explicitly incorporate knowledge of the full table size -- a closed-world luxury.  
Other database techniques include changing the sampling technique to take advantage of blocks in memory, e.g., \cite{block_sampling}, or focus on distinct-value estimation in a single scan of the database \cite{gibbons}. 
In the following we focus on estimators that are suitable for use in the open-world. 


\section{Estimating Completeness}
\label{sec:cardest}
Our goal is to reason about query results by estimating completeness as answers arrive from the crowd. 
As described above, we can apply species estimation techniques in the context of crowdsourced queries by drawing the analogy of estimating cardinality of the query result set.
We first discuss the species estimation problem and describe several estimators that vary in the assumptions placed on the underlying distribution over the items in the result set. 
We then compare their performance on example queries.

\subsection{Uniform Estimators} 
\label{sec:card:uniform}
Receiving answers from workers is analogous to drawing samples from some underlying distribution of unknown size $N$; each answer corresponds to one sample from the item distribution.
We can rephrase the problem as a species estimation problem as follows. 

The set of HITs received from AMT is a sample of size $n$ drawn with replacement from a population\footnote{Actually, workers do not sample with replacement, see Section~\ref{sec:sample_repl}.} in which elements can be from $N$ different classes, numbered $1-N$ ($N$, unknown, is what we seek); $c$ is the number of unique classes seen in the sample. 
Let $n_i$ be the number of elements in the sample that belong to class $i$, with $1\leq i\leq N$. 
Of course some $n_i= 0$ because they have not been observed in the sample. 
Let $p_i$ be the probability that an element from class $i$ is selected by a worker, $\sum_{i=1}^{N} p_i = 1$; such a sample is often described as a multinomial sample \cite{bunge_review93}.

If we initially assume a uniform item distribution, each class is equally likely to be selected: $(p_1 = p_2 = \cdots = p_N)$; this transforms the species estimation problem into a simple inference with the single parameter $N$. 
An approximate maximum likelihood estimator (MLE) is the solution $N$ of the equation \cite{turing53}: 
\begin{equation}
c = N(1-e^{-n/N})
\label{eq:uniform_mle}
\end{equation}

This solution is related to classic urn sampling problems like the coupon collector or occupancy problems \cite{feller,flajolet_ccp}.

\subsection{Non-Uniform Estimators} 
\label{sec:card:skew}
Estimators that assume an underlying uniform distribution often work for item distributions that have low skew, as we show in the next subsection. 
When the item distribution is heavily skewed, however, new unique items are acquired more slowly than in the uniform case. 
Thus the cardinality estimate produced by an estimator assuming equi-probable items will be an underestimate and can be thought of as a lower-bound \cite{bunge_review93}.
In the crowdsourcing regime, non-uniformity occurs when workers are more inclined to respond with some particular answers as compared to others; skew can be inherent in the data or due to how workers find their answers. 
For example, in the US states experiments, the five states that workers tended to provide early on were California, New York, Alabama, and Florida, and Texas\footnote{The four most popular states, as well as the first state alphabetically}. 
Worker answer sequences for the UN experiments often appeared in alphabetical order, sometimes preceded by popular large countries like the US, India, or China. 

To cope with skew, many estimators use a statistic called the ``frequency of frequencies", discussed next. 
Later we describe the estimators that incorporate this metric. 

\subsubsection{Frequency of Frequencies} 
The ``frequency of frequencies" statistic $f$ captures the relative frequency of the observed samples. 
For a population that can be partitioned into $N$ classes (items), and for a given sample of size $n$, $f_j$ is defined as the number of classes that have exactly $j$ members in the sample. 
Notably, $f_1$ represents the ``singletons'' and $f_2$ the ``doubletons''. 

\begin{figure}[t]
	\centering
 	\includegraphics[width=3.5in]{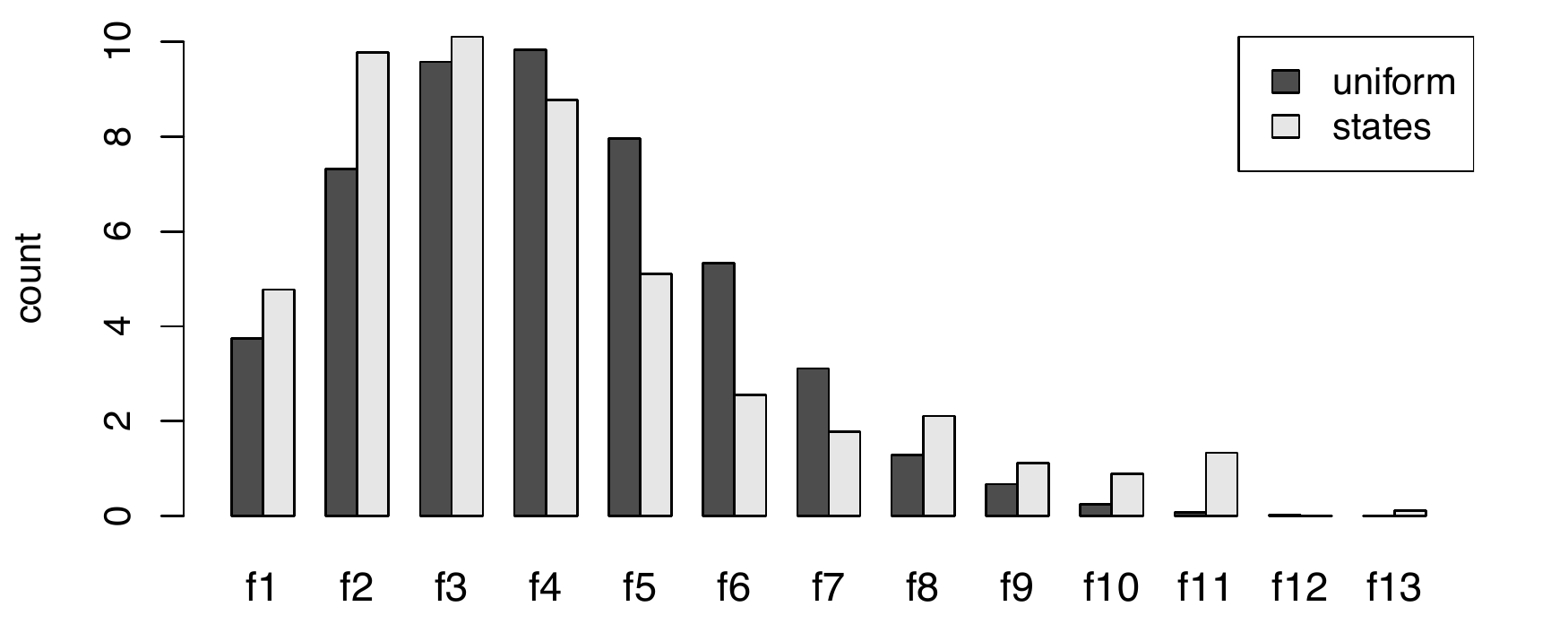}
    \vspace*{-20pt}
	\caption{$f$-statistic histogram after 200 responses.
	}
     \label{fig:fstat}
\vspace*{-10pt}
\end{figure}

To illustrate the effect of skew on the $f$-statistic, Figure~\ref{fig:fstat} shows a histogram for acquiring the 50 US states from the crowd after 200 HITs and compares it to synthetically drawing 200 samples from a uniform distribution over 50 unique classes. 
The bars are averaged over the nine runs of the experiment. 
After 200 samples from a uniform distribution over 50 items, one would expect most items would have appeared approximately four times; indeed the dark bars are bell-shaped centered at $f_4$. 
In contrast, the states experiment has more mass on the higher $f$'s, indicating that some states appear very frequently (popular states like New York and California). 
In general, concentration of mass around one $f_j$ indicates a uniform distribution; more item skew will spread the mass across the $f$'s. 
The intuition behind using the $f$-statistic for estimating the number of total items is that the presence of rare items (e.g., $f_1$) indicates the likely existence of other items that are not currently represented in the dataset. 

\label{sec:fstat}

One might try to estimate the underlying distribution $p_1 ... p_N$ in order to predict the cardinality $N$. 
However, Burnham and Overton \cite{bo78} show that the $f$-statistic is a sufficient statistic for estimating $f_0$, the number of unobserved species.  
Thus the goal is to form a cardinality estimate by predicting the value of $f_0$. 
 
Parametric approaches attempt to predict $f_0$ by fitting an existing distribution to the $f$-statistic, like a lognormal or inverse gaussian.
The problem with the parametric approach is that the estimate will be poor if the chosen distribution does not fit the data well; furthermore the choice of distribution for one use case might not hold for another. 
Non-parametric approaches use only the $f$-statistic, thereby putting no restrictions on the underlying distribution. 
Two common non-parametric estimators are \emph{Chao84} and \emph{Chao92}, described next. 

\subsubsection{Chao84 Estimator}
In \cite{chao84}, Chao develops a simple estimator for species richness that is based solely on the number of rare species found in the sample:
\vspace{-0.1in}
\[
\hat{N}_{chao84} = c + \frac{f_1^2}{2f_2}  
\]
Chao found that it actually is a lower bound, but it performed well on her test data sets. 
She also found that the estimator works best when there are relatively rare species, which is often the case in real species estimation scenarios. 

\begin{figure*}[tb]
  \begin{tabularx}{\textwidth}{C C C}
	\begin{minipage}{.32\textwidth}
      \centering
      (a) Avg. States
\vspace{-5pt}
\includegraphics[width=2.3in]{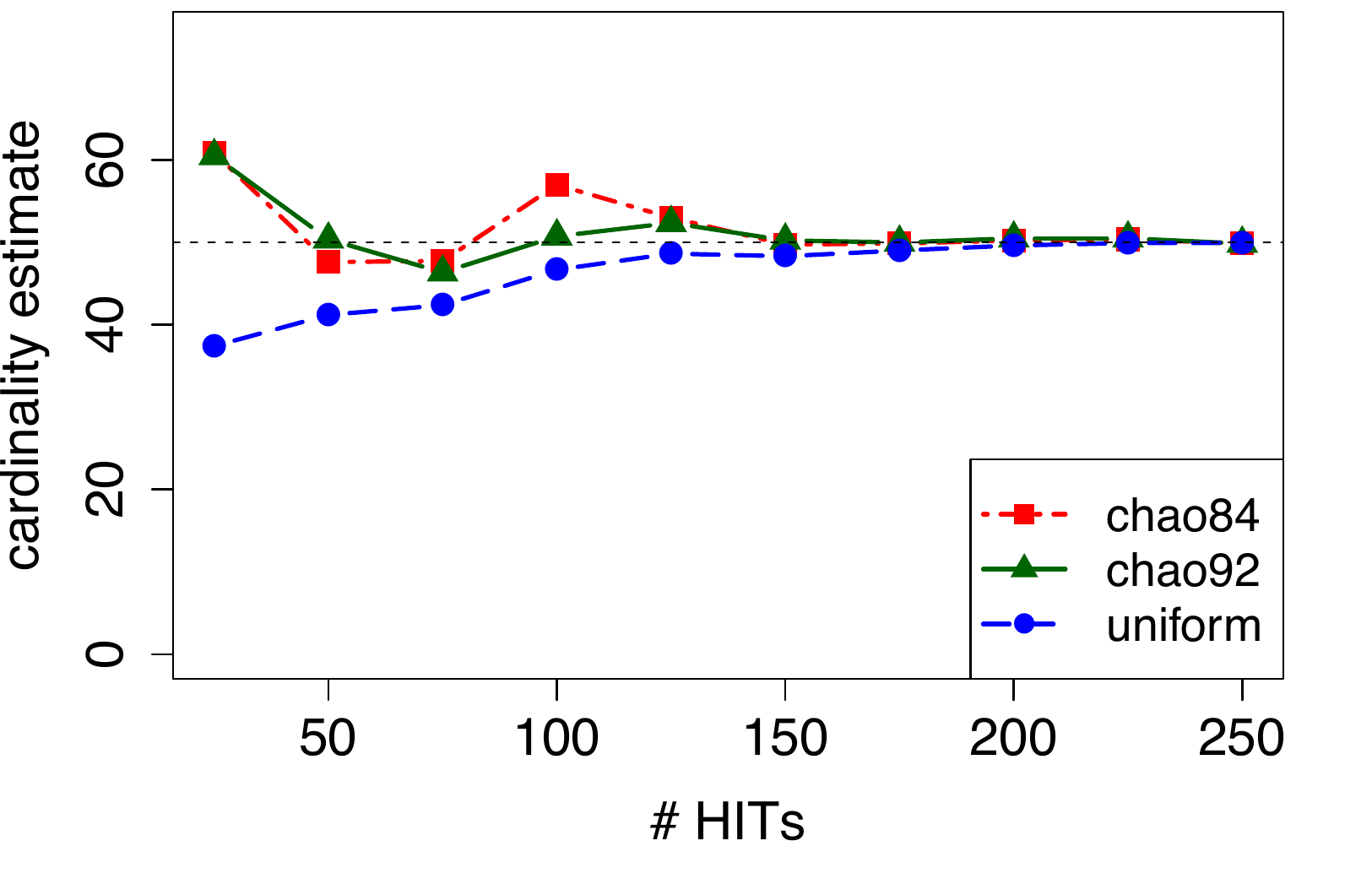}
\includegraphics[width=2.1in]{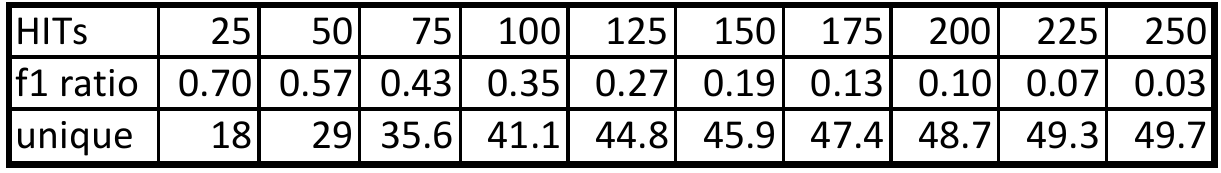}
    \end{minipage} &
	\begin{minipage}{.32\textwidth}
      \centering
	  (b) Avg. UN 
	\vspace{-5pt}
	\includegraphics[width=2.3in]{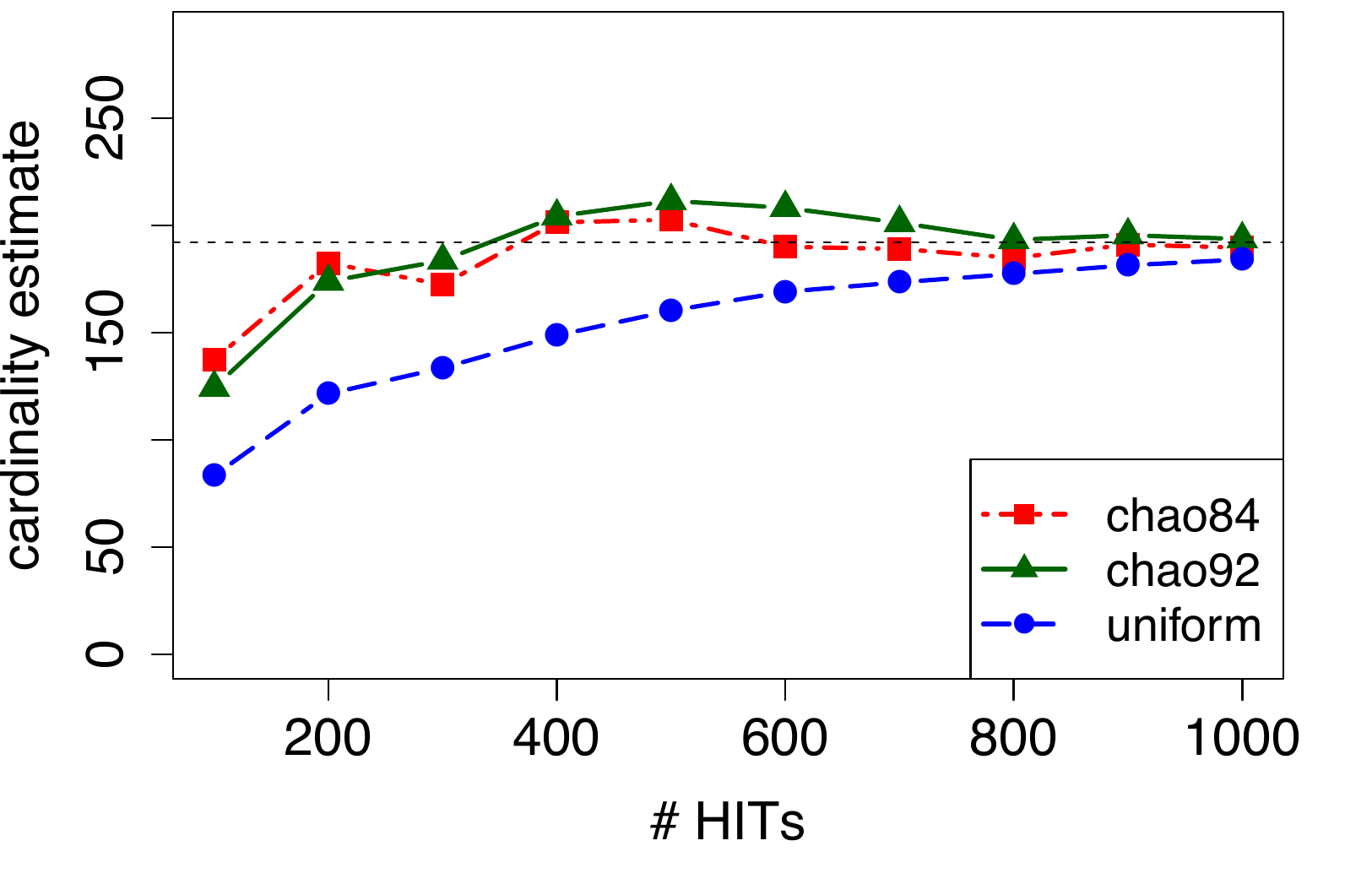}
	\includegraphics[width=2.1in]{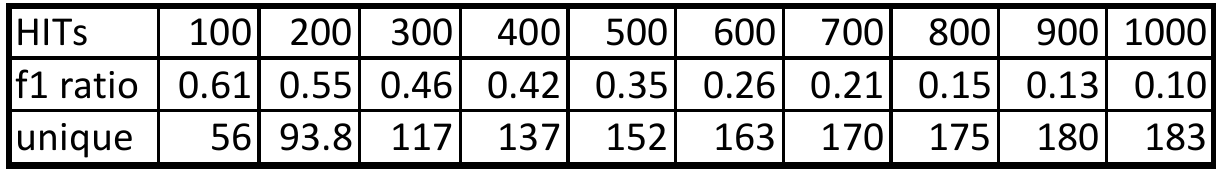}
    \end{minipage}  &
	\begin{minipage}{.32\textwidth}
      \centering
	  (c) Ice Cream 
	\vspace{-5pt}
	\includegraphics[width=2.3in]{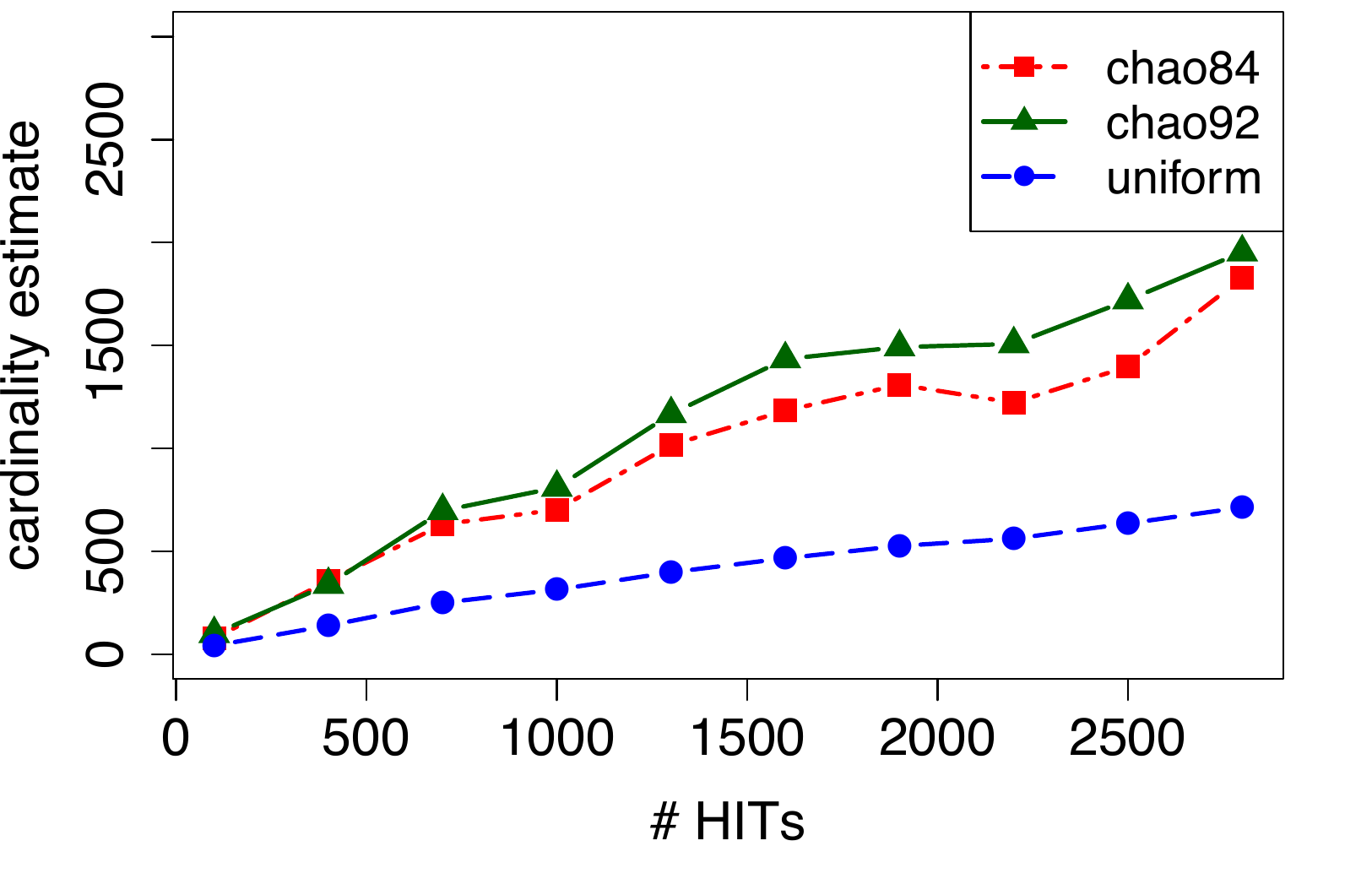} 
	\includegraphics[width=2.1in]{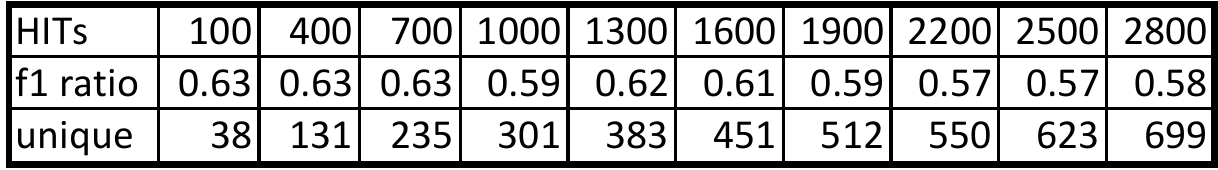} 
    \end{minipage} \\
%
  \end{tabularx}
	\vspace*{-10pt}
	\caption{(a-c) Above: cardinality estimates for different experiments. Below: f1-ratio and number of unique items for increasing numbers of HITs.}
	\label{fig:cardest}
\end{figure*}

\subsubsection{Chao92 Estimator}
\label{sec:chao92}
In \cite{chao92}, Chao develops another estimator based on the notion of \emph{sample coverage}. 
The sample coverage $C$ is the sum of the probabilities $p_i$ of the observed classes. 
However, since the underlying distribution $p_1 ... p_N$ is unknown, this estimate from the Good-Turing estimator\cite{turing53} is used: 
\[
\hat{C} = 1 - f_1/n
\]
The Chao92 estimator attempts to explicitly characterize and incorporate the skew of the underlying distribution using the \emph{coefficient of variance} (CV), denoted $\gamma$, a metric that can be used to describe the variance in a probability distribution \cite{chao92}; we can use the CV to compare the skew of different class distributions.  
The CV is defined as the standard deviation divided by the mean. 
Given the $p_i$'s ($p_1 \cdots p_N$) that describe the probability of the $i$th class being selected, with mean $\bar{p} = \sum_i p_i/N = 1/N$, the CV is expressed as $\gamma = \left[\sum_i(p_i - \bar{p})^2/N\right]^{1/2} \left. \right/ \bar{p}$ \cite{chao92}. 
A higher CV indicates higher variance amongst the $p_i$'s, while a CV of 0 indicates that each item is equally likely.

The true CV cannot be calculated without knowledge of the $p_i$'s, so Chao92 uses an estimate, $\hat{\gamma}$. 
\begin{equation}
\label{eqn:cv_est}
\hat{\gamma}^2 = \max \left\{\frac{c}{\hat{C}} \sum_i{ i(i-1)f_i}  \bigg/  n(n-1) - 1 ,0\right\}
\end{equation}
The estimator that uses the coefficient of variance is below; note that if $\hat{\gamma}^2 = 0$ (i.e., indicating a uniform distribution), the estimator reduces to $c/\hat{C}$
\[
\hat{N}_{chao92} = \frac{c}{\hat{C}} + \frac{n(1-\hat{C})}{\hat{C}}\hat{\gamma}^2  
\]

\subsection{Experimental Results} 
\label{sec:card:eval}
We ran over 25,000 HITs on AMT to compare the performance of the different estimators.  
Several \texttt{CROWD} tables we experimented with include small and large well-defined sets like NBA teams, US states, UN member countries, as well as less well-defined sets like ice cream flavors, animals found in a zoo, and graduate students about to graduate. 
Workers were paid \$0.01 to provide one item in the set using the UI in Figure~\ref{fig:ui}; they were allowed to complete multiple tasks if they wanted to submit more than one answer. 

In the remainder of this paper we focus on three experiments, US states, UN member countries, and ice cream flavors, to demonstrate a range of characteristics that a particular query may have.  
The US states is a small, constrained set while the UN countries is a larger constrained set that is not so readily memorizable. 
The ice cream flavors experiment captures a set that has a fair amount of membership and size ambiguity. 
We repeated our experiments nine times for the US states, five times for the UN countries and once for the ice cream flavors. 
In this paper we cleaned and verified workers' answers manually; other work has described techniques for crowd-based verification \cite{panos_quality, automan, QoE, crowdtutorial}. 

Figure~\ref{fig:cardest}(a-c) shows the average cardinality estimates over time, i.e., for increasing numbers of HITs, for the US states, UN countries, and ice cream flavors using three different estimators. 
Error bars can be computed using variance estimators provided in \cite{chao92,chao84}, however we omit them for better readability. 
The horizontal line indicates the true cardinality if it is known. 
Below each graph, a table shows the ``f1-ratio" and the actual number of received unique items over time.  
We define f1-ratio as $f_1/\sum_i{f_i}$, the fraction of the singletons as compared to the overall received unique items. 
Recall that the presence of singletons is a strong indicator that there are more undetected items; when there are relatively few singletons, we have likely approached the plateau of the SAC. 
The f1-ratio can be used as an indication of whether or not the sample size is sufficient for stable cardinality estimation. 
Since estimators use the relative frequencies of $f_1$ compared to the other $f_i$'s, a high f1-ratio will make it more difficult for the estimators to converge.  
Also note that the ratio between the unique items and the predicted cardinality is the completeness estimate.

\subsubsection{US States}
For the US states (Figure~\ref{fig:cardest}(a)), all estimators perform fairly well; Chao92 remains closer to the true value than Chao84. 
The estimates are stable at 150 HITs, and near the true value even earlier. 
Note this happens well before all fifty states are acquired (on average, after 225 HITs). 
It may be be surprising that the uniform estimator performs as well as it does, as one might suspect that certain states would be more commonly chosen than others. 
There are a few explanations for this performance. 
First, the average coefficient of variance $\hat{\gamma}$ for the states experiments is $0.53$; in \cite{chao92}, Chao notes that the uniform estimator is still reasonable for  $\gamma \leq 0.5$. 
Furthermore, individual workers typically do not submit the same answer multiple times; samples drawn without replacement from a particular distribution will result in a less skewed distribution than the original.  
We discuss sampling without replacement further in Section~\ref{sec:sample_repl}. 
Individual workers also may be drawing from different skewed distributions, e.g., naming the midwestern states before those in the mid-atlantic.

\subsubsection{UN Countries}
In contrast to the US states, the uniform estimator more dramatically under-predicts the true value of 192 for the UN countries experiments (Figure~\ref{fig:cardest}(b)).
This makes sense considering the average $\hat{\gamma}$ for the UN countries experiments is $0.73$. 
Since the uniform estimator assumes each country is equally likely to be given, it predicts that the total set size is smaller. 
The Chao estimators converge faster to the true value than the uniform estimator. 
Unlike the States experiment, we did not obtain the full set in most of the experiment runs (see the table in Figure~\ref{fig:cardest}(b)).

The Chao estimators produced good predictions, however they appeared to fluctuate in the middle of the experiment, starting low then increasing over the true value before converging back down. 
While it is encouraging that the estimators perform well on average, we observed that the variance was fairly high -- indicating to us that some of the experiment runs did not act as expected. 
The classic theory starts to break down in these scenarios because it does not consider crowd-specific behaviors that influence how answers arrive. 
One such behavior is the uneven distribution of the number of answers submitted by participating workers, which can cause estimators to over-predict the cardinality of the result set. 
We address this issue in Section~\ref{sec:sample_repl}. 

\subsubsection{Ice Cream Flavors}
In both the US states and UN countries experiments, the estimators converged and we were able to obtain almost the entire set in the number of posted HITs. 
However, some sets are so large and/or have such skewed item distributions that the cardinality estimate does not converge within the amount of worker answers obtained; this is the case with the ice cream flavors experiment. 
Both the coefficient of variance $\gamma$ and the f1-ratio give insight into this case and allow us to detect it. 
Recall that a very high $\gamma$ value indicates high skew in the item distribution, whereas a high f1-ratio indicates a large set size compared to the current sample size. 
In the ice cream flavors experiment, we found both a high $\hat{\gamma}$ of $5.84$ (compared to 0.53 in the States experiment) and a high f1-ratio that decreases very slowly over time (Figure~\ref{fig:cardest}(c)). 
Both qualities contribute to the estimator's lack of convergence: if we are still receiving many new items, we have not reached the plateau of the SAC. 
Furthermore, estimators tend to under-predict cardinality for very high skew because there is always a chance to see more items from the long tail of the item distribution. 
This suggests we might have to think differently about how to reason about queries over such sets.

\subsection{Discussion}
\label{sec:cardest_disc}
The US States and UN countries experiments show that species estimation techniques, particularly the algorithms that target skew in the item distribution, are able on average to predict the cardinality of a crowdsourced set with manageable skew and size. 
In most of these cases, the analogy between a stream of workers' answers and samples drawn from some distribution is effective. 
In some cases, as with several UN experiment runs, we saw that the crowd exhibits unique behavior that chips away at the classic theory's assumptions. 
We discuss in Section~\ref{sec:sample_repl} how an uneven distribution of the number of answers from workers can cause the estimator to over-predict and provide heuristics that can compensate for that effect. 
Another subtle crowd behavior we observed is workers getting their answers from the same lists found on the web; we defer this discussion and our detection heuristic to Section~\ref{sec:listwalking} as the behavior does not influence the estimators. 

There may be some instances where the number of workers' answers is not sufficient for the estimator to converge due to set size and high skew; the ice cream experiment is an example of such a case. 
However, in such scenarios predicting the total set size does not make sense, an observation that has also been made in the context of species estimation. 
Good, who worked on this problem with Turing, stated in 1953: ``I don't believe it is usually possible to estimate the number of species... but only an appropriate lower bound for that number. This is because there is nearly always a good chance that there are a very large number of extremely rare species'' \cite{bunge_review93}.
With too few answers for a set size prediction and/or a highly skewed item distribution, a more appropriate way to reason about the query result is through the cost-benefit analysis of expending additional effort. 
At some point, the cost of further set enumeration exceeds its usefulness to the user. 
We discuss the notion of \emph{pay as you go} in Section~\ref{sec:payg}. 
 
In the following sections, we describe and propose new techniques to address the crowd-specific issues that impact cardinality estimation, as well as provide techniques to reason about the cost vs. benefit tradeoff of ``getting it all''. 
For the rest of the paper, we use the Chao92 estimator because it provides good overall performance independent of the underlying distribution and has less variance than Chao84. 


\section{Workers and Estimators}
\label{sec:sample_repl}
Species estimation techniques provide a viable foundation for the goal of estimating query progress as answers arrive. 
However, sometimes crowd-specific behaviors can impact the estimator. 
This can happen because the answers that human workers provide are different than simple with-replacement samples. 
Most of the time, workers do not provide the same answer twice.\footnote{Workers may not do bad work that is verifiably against what the requester intended \cite{mason_honesty}}
In other words, an individual worker is sampling \emph{without} replacement from the item distribution. 
Also, often a few overzealous workers provide the majority of answers; i.e., the distribution of answers from workers is skewed. 
In this section, we show that worker skew exists in our AMT experiments and give a heuristic to correct the sampling bias introduced by skew. 

\begin{figure}[t]
	\begin{center}
	\includegraphics[width=2.8in]{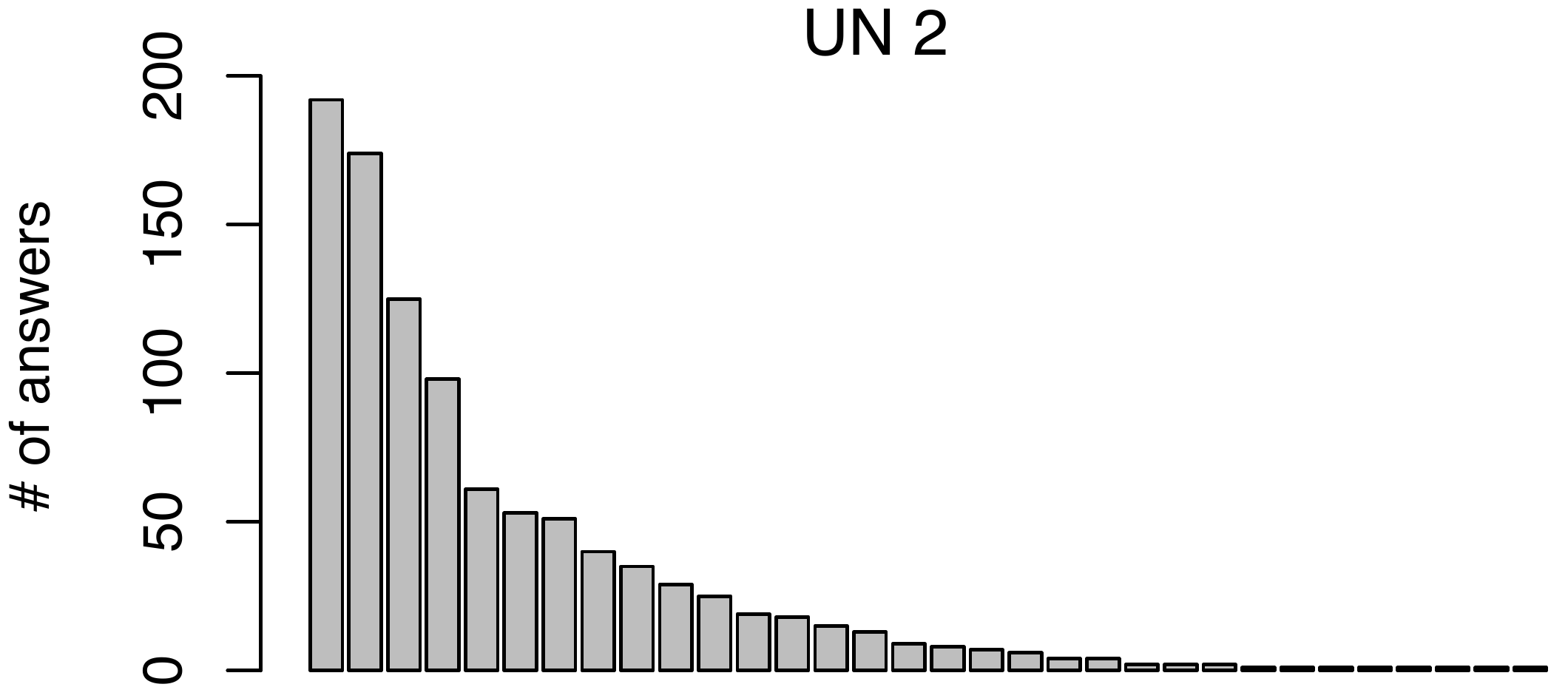}
	\vspace*{-20pt}
	\end{center}
	\caption{Worker distribution of answers for one of the UN experiment runs.}
	\label{fig:streakers}
	\vspace*{-10pt}
\end{figure}

\begin{figure*}[tb]
  \begin{tabularx}{\textwidth}{c c c}
	\begin{minipage}{5.5cm}
 \centering
      (a) Avg. UN
		\includegraphics[width=5.5cm] {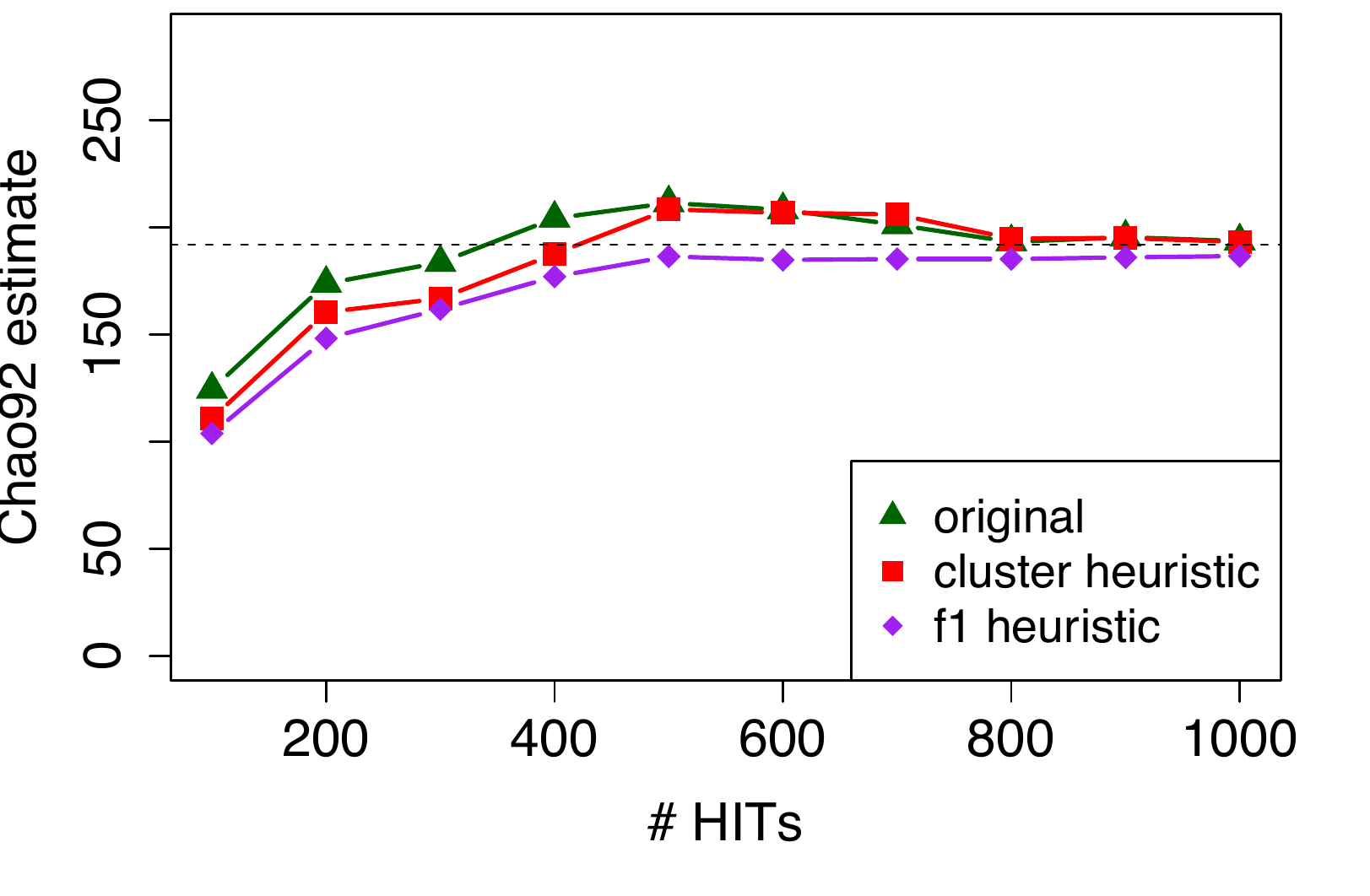}
    \end{minipage} &
	\begin{minipage}{5.5cm}
		 \centering
	  (b) UN2
	\includegraphics[width=5.5cm] {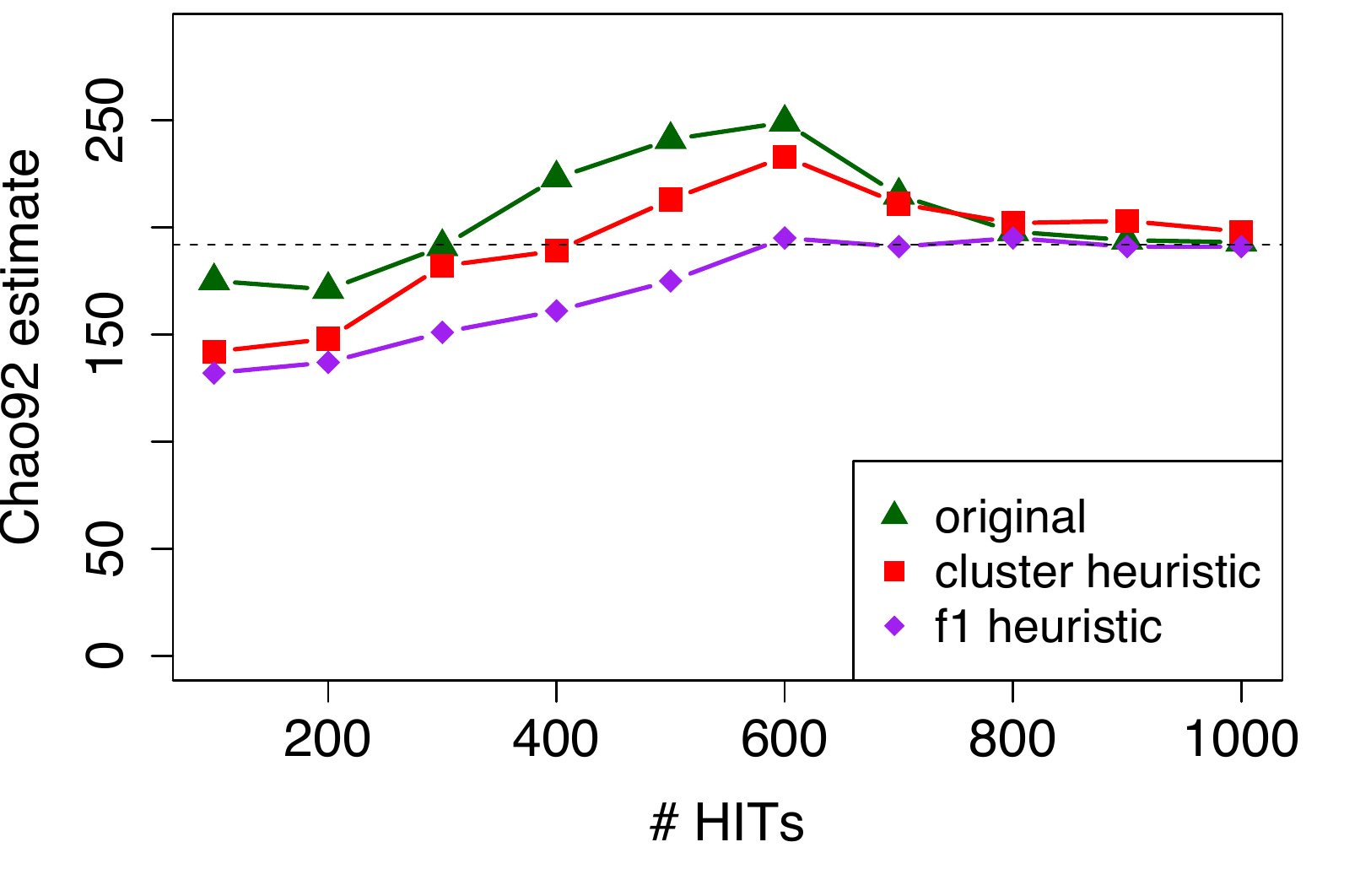}	
    \end{minipage}  &
	\begin{minipage}{5.5cm}
		 \centering
    	(c) UN3
	\includegraphics[width=5.5cm] {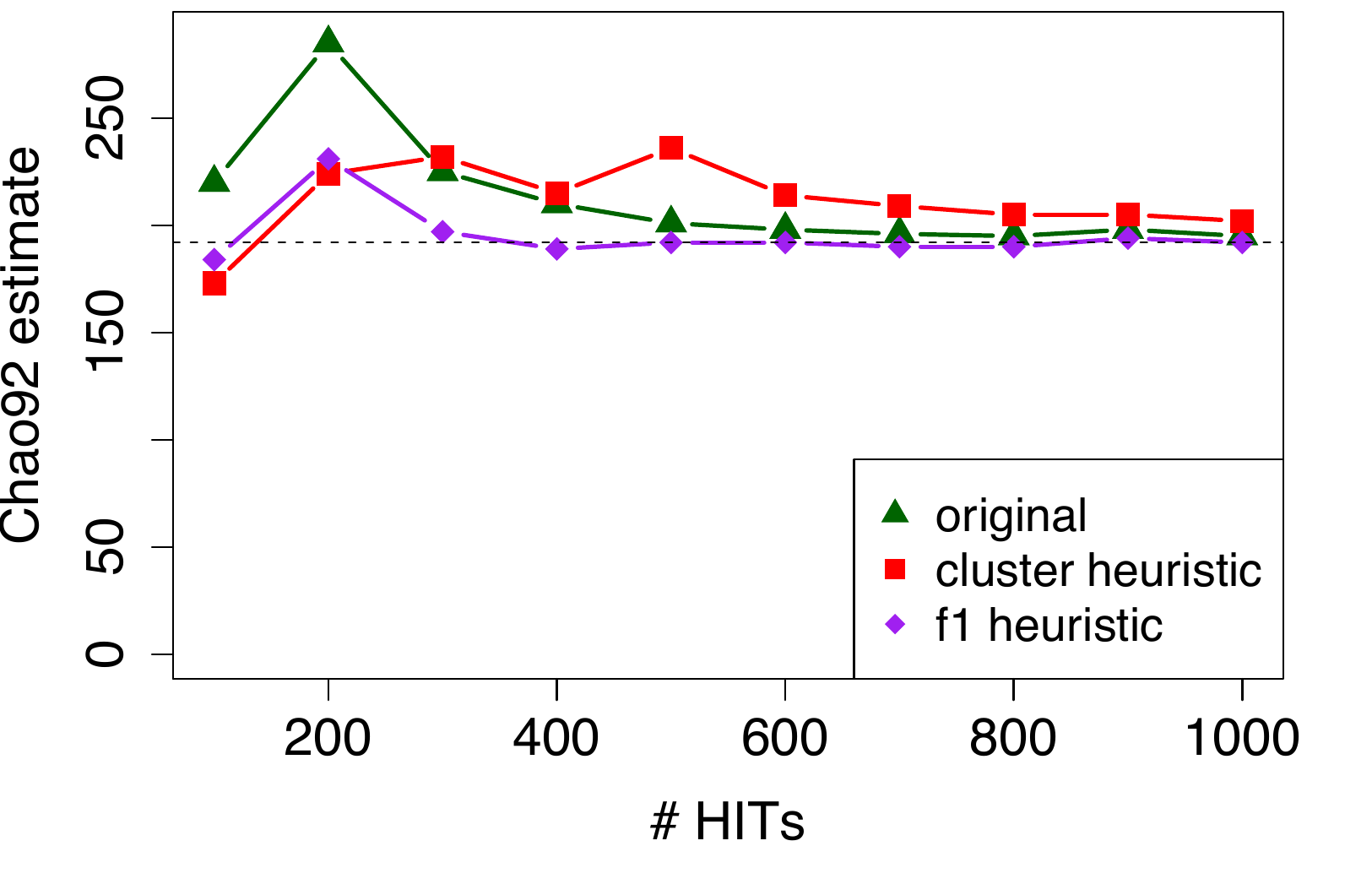}	
    \end{minipage}
  \end{tabularx}
	\vspace*{-10pt}
	\caption{(a) Heuristics applied all UN runs, averaged (b-c) Heuristics applied to UN 2 and 3}
	\label{fig:sampling}
	\vspace*{-10pt}
\end{figure*}

\subsection{Streakers vs. Samplers}
Recall that the Chao92 estimator is heavily informed by the presence of singleton ($f_1$) answers present in the sample. 
When individual workers sample without replacement, new unique items can appear more quickly than expected. 
Imagine the extreme case: a single worker provides only singleton answers, yielding an infinite cardinality estimate. 
In contrast, if each answer comes from a different worker, the resulting sample would be in concordance with a with-replacement sample.
In a simulation varying the number of workers between these two extremes, we found minimal impact on the estimator with more than 9 or 10 workers. 

On crowdsourcing platforms like AMT it is common that some workers complete many more HITs than others. 
This skew in relative worker HIT completion has been labeled the ``streakers vs. samplers'' effect \cite{streakers}. 
The streakers are those workers who really enjoy the task and/or want to amortize the time spent learning how the tasks works by doing many of them. 
Samplers, on the other hand, only try a few tasks or do not have enough time to do more than a few.
The impact of worker skew on cardinality estimation is similar to having too few total workers: 
since they sample without replacement, streakers provide many unique answers that dominate the sample, causing the estimator to over-predict. 

We observed worker skew in our experiments as well, for example Figure~\ref{fig:streakers} depicts the distribution of answers per worker in one of the UN experiment runs. 
Each bar represents the number of answers from an individual worker for the entire experiment run.  
Also note that workers both start and stop providing items at different times during the experiment. 
At any point in time, streakers may be more or less prevalent and their impact may not only be visible at the beginning of the experiment. 
The appearance of streakers at various times during an experiment run influenced the estimator's performance in several UN experiment runs, but had little effect on the US states experiments.


\subsection{Reducing Streaker Impact}
\label{sec:samp_heuristic}
In the following, we propose heuristics for reducing the impact that worker skew has on the Chao92 estimator. 
The intuition behind the heuristics is to ``slow down'' overzealous workers by limiting their contribution to the evaluation of the estimator. 
One heuristic is a simple truncation to reduce the influence of the top workers, while the second one extends it to target only new unique items using knowledge of the $f$-statistic. 

\subsubsection{Multistage Cluster Heuristic}
The multistage cluster heuristic is so named because it is inspired by multistage cluster sampling \cite{sampling_techniques}, in which samples are drawn from a population in stages. 
The first stage of sampling is done by the workers: their answers are samples drawn from the item distribution. 
For the second stage, we sample from each worker's answers and thereby limit the contribution from the top workers. 

More explicitly, in the heuristic we limit the number of answers from any particular worker from exceeding a quota $q$. 
Before evaluating the cardinality estimator, we transform the input by truncating the answer sequence from a worker that has more than $q$ answers. 
We can define $q$ as the average number of answers provided by the top $t$ workers. 
In other words, if the $j$th worker has $a_j$ answers, we remove $max(a_j - q,0)$ of those answers when computing the estimate. 
However, we also want to prevent reducing the sample size too dramatically, as this will decrease the accuracy of the estimator. 
Thus we remove no more than $r$\% of a worker's answers. 
Higher values of $r$ will make the streakers' contributions more balanced, but drastically reducing the sample size will also impact the estimator's accuracy. 
Of course, there is a trade-off in the choices of $t$ and $r$; higher values for both will decrease the sample size, particularly if the samplers produce very few answers; we use $t=10$ based on the simulation mentioned above and $r=40\%$.  
After determining how many answers to retain from a particular worker (which is at most $q$), we sample without replacement that many answers from the worker's original answer sequence.

\subsubsection{f1-Heuristic}
The previous heuristic does not distinguish between the singletons and the answers that fall into the other $f_j$. 
Our goal is to prevent estimator over-prediction due to rapid appearance of \emph{new} items; thus we should target the answers that are part of the $f_1$ set. 
Truncating doubleton, tripleton, etc. responses may actually increase the number of singletons because we may be removing duplicates, potentially causing the estimator to over-predict again.

We amend the previous heuristic to reduce only the number of singletons that streakers contribute. 
Now let $a_j$ be the number of answers from the $j$th worker that are in the set of $f_1$ answers. 
We set $q$ to be the average number of $f_1$s provided by the top $t$ workers. 
We remove $max(a_j - q,0)$ of the $f_1$ answers before computing the estimate (but not more than $r\%$, as before). 
Both heuristics will behave similarly when streakers contribute mostly singleton answers. 
However, when the appearance of new items wanes, the f1-heuristic will remove few, if any, answers. 

\subsection{Experimental Results}
Figure~\ref{fig:sampling}(a) shows the original Chao92 estimates as well as the estimates after the two heuristics have been applied for the averaged UN experiments. 
We additionally highlight two runs in particular with pronounced streaker issues that influenced the cardinality prediction. 
The f1-heuristic converges faster on average, but does not look dramatic because the heuristic has little effect on the estimator if there is little or no streaker issue. 
The impact is more visible in specific runs. 
Figures~\ref{fig:sampling}(b) and (c) depict two examples where the heuristics had significant impact. 
In both cases, the f1-heuristic greatly reduces the over-prediction bumps seen in the original Chao92 estimate; in the latter case, the restriction $r$ on the amount of data the heuristic can exclude results in a small over-prediction in the beginning. 


For both heuristics, the impact of the streakers is visibly lessened towards the beginning of the experiment. 
However, the cluster heuristic tends to over-predict again later on. 
As previously mentioned, this likely happens because excluding answers from the streakers can also be removing duplicates or triplicates, which the estimator interprets as the presence of more unseen items. 
In contrast, the f1-heuristic ensures that we only target new items that the streakers introduce by taking into consideration the impact of truncating on the $f$-statistic which influences the estimator. 
The heuristic works well for reducing the impact of streakers and making the sample more reasonable for the estimator despite sampling without replacement.


\section{Cost vs. benefit: pay-as-you-go}
\label{sec:payg}
The algorithms developed for species estimation work well on average for predicting the query result set size in the US States and UN countries experiments, and we have shown heuristics to remedy the crowd-specific behaviors in particular runs. 
However, recall that the estimators in the ice cream experiment were not able to converge in the number of answers obtained from the crowd. 
As we discussed in Section~\ref{sec:cardest_disc}, the result set for many reasonable queries may have unbounded size and/or a highly skewed distribution that make predicting its size nonsensical. 
For these types of queries, it makes more sense to try to estimate the benefit of spending more money, i.e., predicting the shape of the SAC in the near future. 
Eventually, the cost of getting a few more answers is prohibitively expensive or impossible and thus it makes sense to \emph{pay as you go}. 


In this section, we apply several techniques from the species estimation literature for estimating benefit of increased effort. 
We then evaluate these techniques on our example use cases, finding that they perform well considering the different context of crowd-supplied answers. 

\subsection{Estimating Benefit}
An open-world system would want to estimate the benefit of increased crowdsourcing effort in order to consider the end user's goals and incorporate this knowledge into query optimization.  
For the SELECT query in CrowdDB, we are particularly interested in how many more unique items would be acquired with $m$ more HITs, for a given number of current HITs. 
In the following, we describe and apply two methods from the species estimation literature to build a pay-as-you-go technique for crowdsourced data. 


\subsubsection{Extrapolating the species accumulation curve}
Recall that the \emph{species accumulation curve} (SAC) depicts the number of unique elements as more worker answers are received. 
A natural approach would be to extrapolate the SAC to see the advantage of posting more HITs. 
For example, if we have observed 34 unique items after receiving 50 worker responses, we would like to estimate how many more unique items we would see if we issued another 50 HITs.
In this paper, we evaluate the {\em spline} technique for extrapolating the curve as described in \cite{colwell_prediction}. 
We first calculate the ``mean'' SAC by permuting the data many times and averaging the SACs from each permutation. 
Afterwards, a cubic spline is fit to this smoothed version of the curve, which in turn is used for the final prediction.

\subsubsection{Sample coverage approach}
In \cite{shen_prediction}, the authors derive an estimator (in the following referred to as {\em Shen}) for the expected number of species $\hat{N}_{Shen}$ that would be found in an increased sample of size $m$. 
It incorporates the notion of the sample coverage $C$ (see Section~\ref{sec:chao92}), and the intuition that $1-C$ is the conditional probability of discovering a new species in a larger sample. 
The approach assumes we have an estimate of the number of unobserved elements $w$ (same as $f_0$) and that the unobserved elements have equal relative abundances. 
However, this cardinality estimate $w$ can incorporate a coefficient of variance estimate (equation \ref{eqn:cv_est}) to account for skew. 
Thus, an estimate of the unique elements found in an increased effort of size $m$ is: 
\vspace{-0.1in}
\begin{equation}
\hat{N}_{Shen} = \hat{w} \left[ 1 - \left(  1 - \frac{1-\hat{C}}{\hat{w}} \right)^m \right]
\label{eqn:chao_payg}
\end{equation}

Another technique \cite{colwell_prediction}  models the ``expected mean'' SAC with a binomial mixture model. It performs similar to the coverage approach; we do not discuss it further. 

\begin{table*}[h]
\begin{center}\scriptsize 
	\begin{tabular}{|c ||c |c |c ||c |c |c ||c |c |c |} \hline
		\multicolumn{10}{|c |} {\bf Average of states experiments}  \\ \hline
		& \multicolumn{3}{c||}{HITs 50} & \multicolumn{3}{c||}{HITs 150} & \multicolumn{3}{c|}{HITs 250} \\
	 	m & actual & Shen & spline & actual & Shen & spline & actual & Shen &  spline \\
	 	10 & \phantom{0}3.38 & \phantom{0}2.99 	& \phantom{0}3.19 	& 0.75 & 0.56 & 0.44 & 0.00 & 0.06 & 0.05 \\
	 	20 & \phantom{0}5.62 & \phantom{0}5.52 	& \phantom{0}6.39 	& 1.00 & 1.05 & 0.88 & 0.13 & 0.11 & 0.11 \\
	 	50 & 12.40 			& 11.00 			& 16.00				& 2.62 & 2.20 & 2.20 & 0.25 & 0.24 & 0.30 \\
	 	100 & 17.50 		& 16.00 			& 31.90 			& 3.50 & 3.38 & 4.41 & 0.25 & 0.38 & 0.60 \\
	 	200 & 21.00 		& 19.60 			& 63.90 			& 3.75 & 4.38 & 8.81 & 0.25 & 0.49 & 1.18 \\ \hline
	\end{tabular}
	\begin{tabular}{|c ||c |c |c ||c |c |c ||c |c |c |} \hline
		\multicolumn{10}{|c |} {\bf Average of the UN experiments}  \\ \hline
		& \multicolumn{3}{c||}{HITs 200} & \multicolumn{3}{c||}{HITs 600} & \multicolumn{3}{c|}{HITs 800} \\
	 m & actual & Shen & spline & actual & Shen & spline & actual & Shen & spline \\
	 10 & \phantom{0}2.20 	& \phantom{0}2.61 	& \phantom{0}2.76 	& \phantom{0}0.80 	& \phantom{0}0.72 	& \phantom{0}0.64 & 0.40 & 0.31 & 0.40 \\
	 20 & \phantom{0}4.60 	& \phantom{0}5.14 	& \phantom{0}5.53 	& \phantom{0}1.40 	& \phantom{0}1.42 	& \phantom{0}1.27 & 0.80 & 0.62 & 0.81 \\
	 50 & 11.20 			& 12.30 			& 13.80 			& \phantom{0}3.60 	& \phantom{0}3.46 	& \phantom{0}3.16 & 2.40 & 1.51 & 2.04 \\
	 100 & 22.80 			& 22.70 			& 27.70 			& \phantom{0}6.60 	& \phantom{0}6.64 	& \phantom{0}6.31 & 4.00 & 2.89 & 4.08 \\
	 200 & 42.80 			& 39.40 			& 55.30 			& 11.80				& 12.30 			& 12.60 & 7.60 & 5.33 & 8.17 \\
	\end{tabular}
	\begin{tabular}{|c ||c |c |c ||c |c |c ||c |c |c |} \hline
		\multicolumn{10}{|c |} {\bf Ice cream experiment}  \\ \hline
		& \multicolumn{3}{c||}{HITs 1000} & \multicolumn{3}{c||}{HITs 1500} & \multicolumn{3}{c|}{HITs 2000} \\
	 m & actual & Shen & spline & actual & Shen & spline & actual & Shen & spline \\
	 10 & \phantom{0}5.00 	& \phantom{0}1.79 	& \phantom{0}1.39 	& \phantom{0}1.00 	& \phantom{0}1.79 	& \phantom{0}1.62 		& \phantom{0}1.00 & \phantom{0}1.54 & \phantom{0}1.25 \\
	 20 & \phantom{0}7.00 	& \phantom{0}3.57 	& \phantom{0}2.77 	& \phantom{0}2.00 	& \phantom{0}3.57 	& \phantom{0}3.27 		& \phantom{0}3.00 & \phantom{0}3.08 & \phantom{0}2.54 \\
	 50 & 17.00 			& \phantom{0}8.91 	& \phantom{0}6.91 	& \phantom{0}7.00 	& \phantom{0}8.91 	& \phantom{0}8.22 		& \phantom{0}7.00	& \phantom{0}7.69 & \phantom{0}6.42 \\
	 100 & 30.00 			& 17.80 			& 13.80 			& 18.00 			& 17.80 			& 16.50 				& 12.00 			& 15.30 & 12.90 \\
	 200 & 55.00 			& 35.20 			& 27.60 			& 39.00 			& 35.40 			& 32.90 				& 23.00				& 30.60 & 25.80 \\ \hline
	\end{tabular}
\caption{Pay-as-you-go: estimation of additional unique items after $m$ more HITs}
\label{table:payg}
\end{center}
\vspace*{-15pt}
\end{table*}

\subsection{Experimental Results}
We evaluated the different pay-as-you-go estimators using the three use cases, US states, UN countries, and the ice cream flavors; we average over the experiments' runs. 
Table~\ref{table:payg} shows the estimates evaluated at different points in time in the experiments (i.e., the current number of received HITs) with varying sizes $m$.
It compares the estimates to the actual number of received unique items after $m$ HITs for both the {\em spline} and {\em Shen} estimator. 
For example in the US states experiment, after having received 150 HITs, the predicted number of additional unique items after posting $m=100$ more HITs is 3.38 with the Shen estimator and 4.41 items with the Spline estimator, whereas the actual number of additional received unique items on average was 3.5.

Both pay-as-you-go estimators are fairly accurate. 
In general, predictions for small $m$ are easier since only the near future is considered. 
The larger the $m$, the further the prediction has to reach and thus the more error-prone the result, particularly if $m$ exceeds the current HITs size \cite{shen_prediction}).

The Shen technique works especially well when there is a lower number of received HITs (i.e., the lower part of the SAC), whereas the spline estimator works slightly better towards the end  after receiving a large number of HITs.
By incorporating the cardinality estimate and coefficient of variance, the Shen estimator reasons about the expected shape of the SAC. 
In contrast, the spline estimator learns the shape of the curve only through the observed samples and has no knowledge about the expected behavior.
This causes the Shen estimator to outperform the spline technique with small samples as it better considers the expected behavior.
However, with a large enough sample size, the spline technique is able to better predict the curve as it has no built-in assumptions such as sampling without replacement.
We also experimented using the Shen technique together with our heuristic from Section~\ref{sec:samp_heuristic}.
Although the technique improves the cardinality prediction, it tends to cause the Shen estimator to under-predict.
This happens because we designed the heuristic to reduce the impact of streakers, which can hide the arrival rate of new items. 

In general the results are aligned with the intuition the SAC provides. 
At the beginning when there are few worker answers, it is fairly inexpensive to acquire new unique items. 
Towards the end, more unique items are hard to come by and, furthermore, the difference in gains between $m=100$ and $m=200$ grows smaller as we enter the plateau of the curve. 
So while the task of``getting them all'' may not make sense in the open-world, asking the question of when there will be diminishing returns allows the system to reason about the quality of the query result.

\newcommand{\mf}{\ensuremath{r}}
\section{List Walking}
\label{sec:listwalking}
When we analyzed the experimental results, we noticed that workers sometimes submit answers in the same order, likely because they consult lists on the web. 
We refer to this effect as {\em list walking}. 
Although not surprising for the UN or States experiments, we were surprised to find list walking even in the ice cream flavor experiment. 
Since list walking can be seen as sampling from a heavily skewed distribution, it can cause the estimators to under-predict and reduce the accuracy of the completeness estimate. 
In theory this could be a problem, however the effect on our experiments was only minor for several reasons. 
Workers used different sources and/or different strategies to provide answers (e.g., starting in the middle of the list, skipping around the list, etc.); this behavior mitigates the impact of list walking.  
Nevertheless, we want to determine the prevalence of list walking to see how much the estimator is affected. 

Furthermore, detecting list walking makes it possible to change the crowd-sourcing strategy. 
For example, we could apply automatic extraction by asking workers for a source URL, or using web browser plugins to scrape the data. 
In cases where one or two lists containing the full set exists, such as the UN countries, this switch could be helpful for getting them all. 
However, it might be harmful to switch strategies for sets for which no single list exists (e.g., ice cream flavors). 

In this section we devise a technique for detecting list walking based on the likelihood that multiple workers provide answers in the same exact order. 
We show that our technique is able to detect and reason about various amounts of list walking in several experiments, including lists that do not appear in alphabetical order.

\subsection{Detecting lists}
The goal of detecting list walking is to differentiate between samples drawn from a skewed item distribution and the existence of a list, which leads to a deterministic answer sequence. 
Simple approaches, such as looking for alphabetical order, finding sequences with high rank correlation or small edit-distance would either fail to detect non-alphabetical orders or disregard the case where workers return the same order simply by chance.

In the rest of this section, we focus on a heuristic to determine the likelihood that a given number of workers $w$ would respond with $s$ answers in the exact same order. 
List walking is similar to extreme skew in the item distribution;
however even under the most skewed distribution, at some point (i.e., large $w$ or large $s$), providing the exact same sequence of answers will be highly unlikely.
Our heuristic determines the probability that multiple workers would give the same answer order if they were really sampling from the same item distribution.
Once this probability drops below a particular threshold (we use $0.01$), we conclude that list walking is likely to be present.
We also consider cases of list walking with different offsets (i.e., both workers started from the fifth item), but we do not consider approximate matches that may happen if workers skip some items on the list. Detecting list use in those scenarios is future work. Furthermore, answer orders that match approximately may make the sample more random and desirable for estimation.

\subsubsection{Preliminary setup: binomial distribution}
Let $W$ be the total number of workers who have provided answer sequences of length $s$ or more. Among these, let $w$ be the number of workers who have the same sequence of answers with length $s$ starting at the same offset $o$ in common. We refer to this sequence as the {\em target sequence} $\alpha$  of length $s$, which itself is composed of the individual answers $\alpha_i$ at every position $i$ starting with offset $o$ ($\alpha=(\alpha_{o+1},\dots,\alpha_{o+s})$). If $p_\alpha$ is the probability of observing that sequence from some worker, we are interested in the probability that $w$ out of $W$ total workers would have that sequence.
This probability can be expressed using the binomial distribution: $W$ corresponds to the number of trials and $w$ represents the number of successes, with  probability mass function (PMF):
\vspace{-0.08in}
\begin{equation}
	Pr(w;W,p_\alpha) = {W \choose w} p_\alpha^w (1-p_\alpha)^{W - w}
\label{eqn:listprob}
\end{equation}
\vspace{-0.02in}
Note that the combinatorial factor captures the likelihood of having $w$ workers sharing the given sequence by chance just because there are many workers $W$.
In our scenario, we do not necessarily care about the probability of exactly $w$ workers providing the same sequence, but rather the probability of \textit{$w$ or more} workers with the same answer sequence:
\vspace{-0.15in}
\begin{equation}
	Pr_{\ge}(w;W,p_\alpha) = 1 - \sum\limits_{i=0}^{w-1} {W \choose i} p_\alpha^i (1-p_\alpha)^{W - i}
\label{eqn:listsum}
\end{equation}

\begin{figure*}[t]
  \begin{tabularx}{\textwidth}{C C C}
    \includegraphics[width=.33\textwidth]{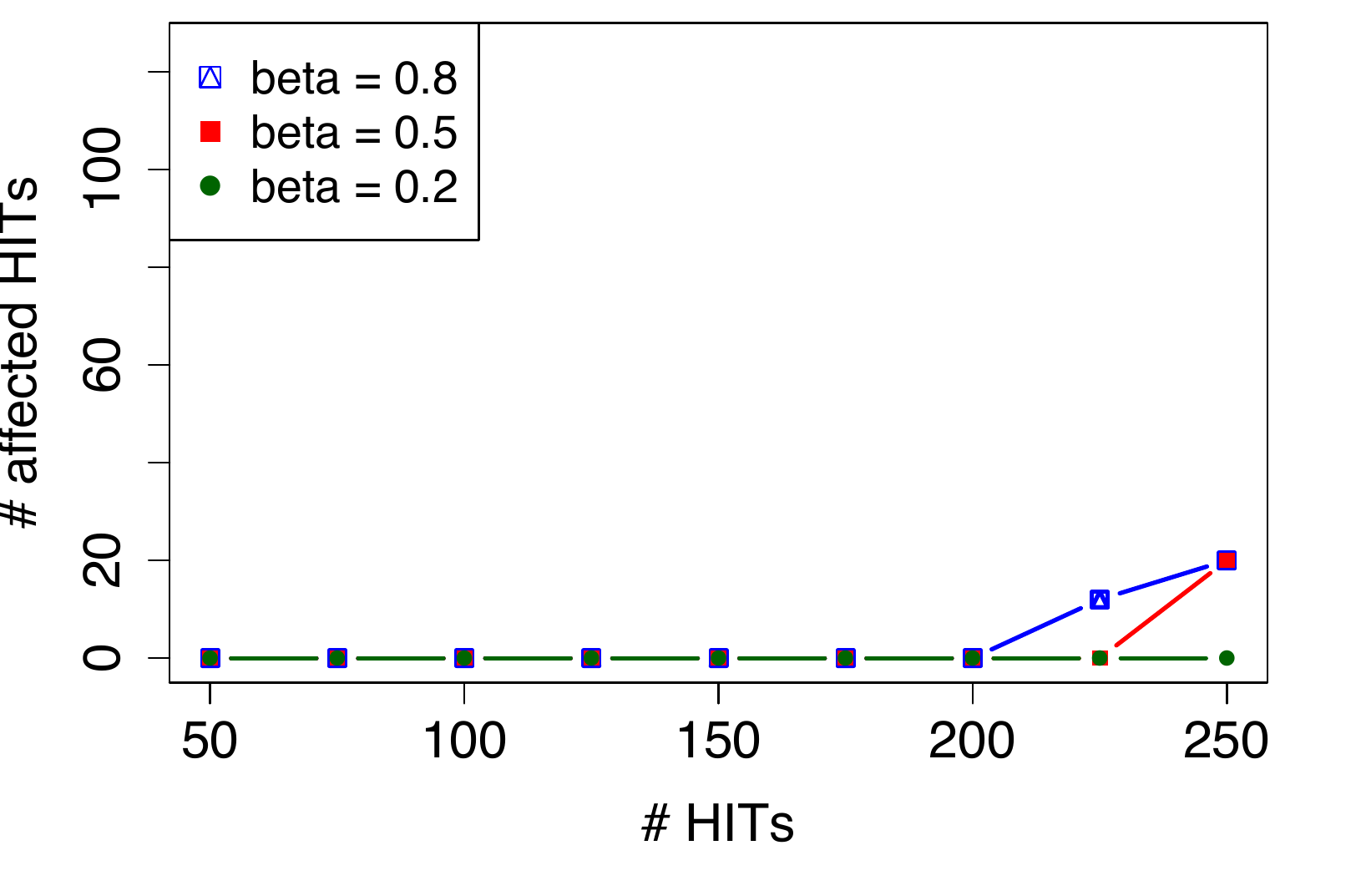} &
    \includegraphics[width=.33\textwidth]{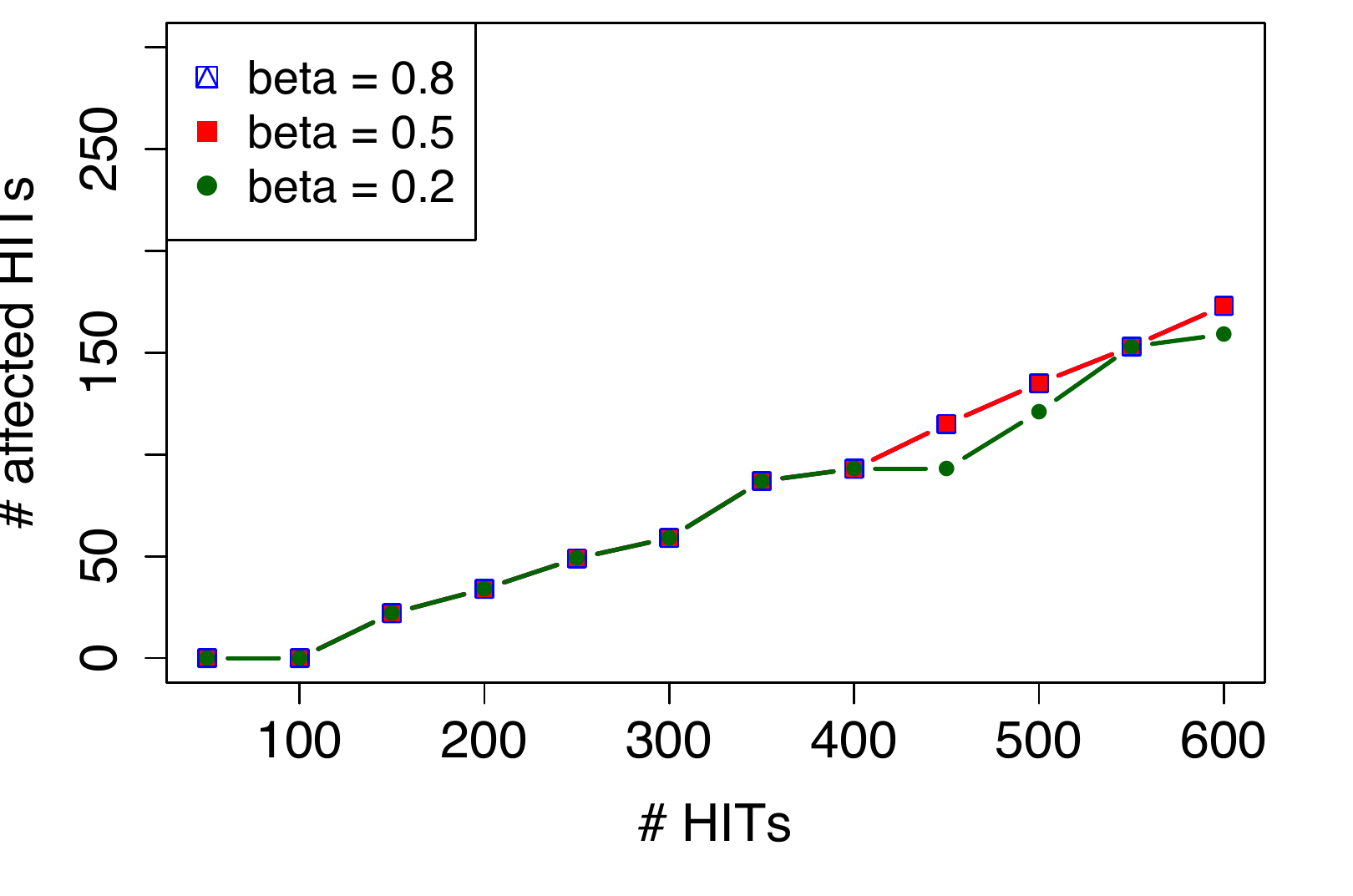} &
    \includegraphics[width=.33\textwidth]{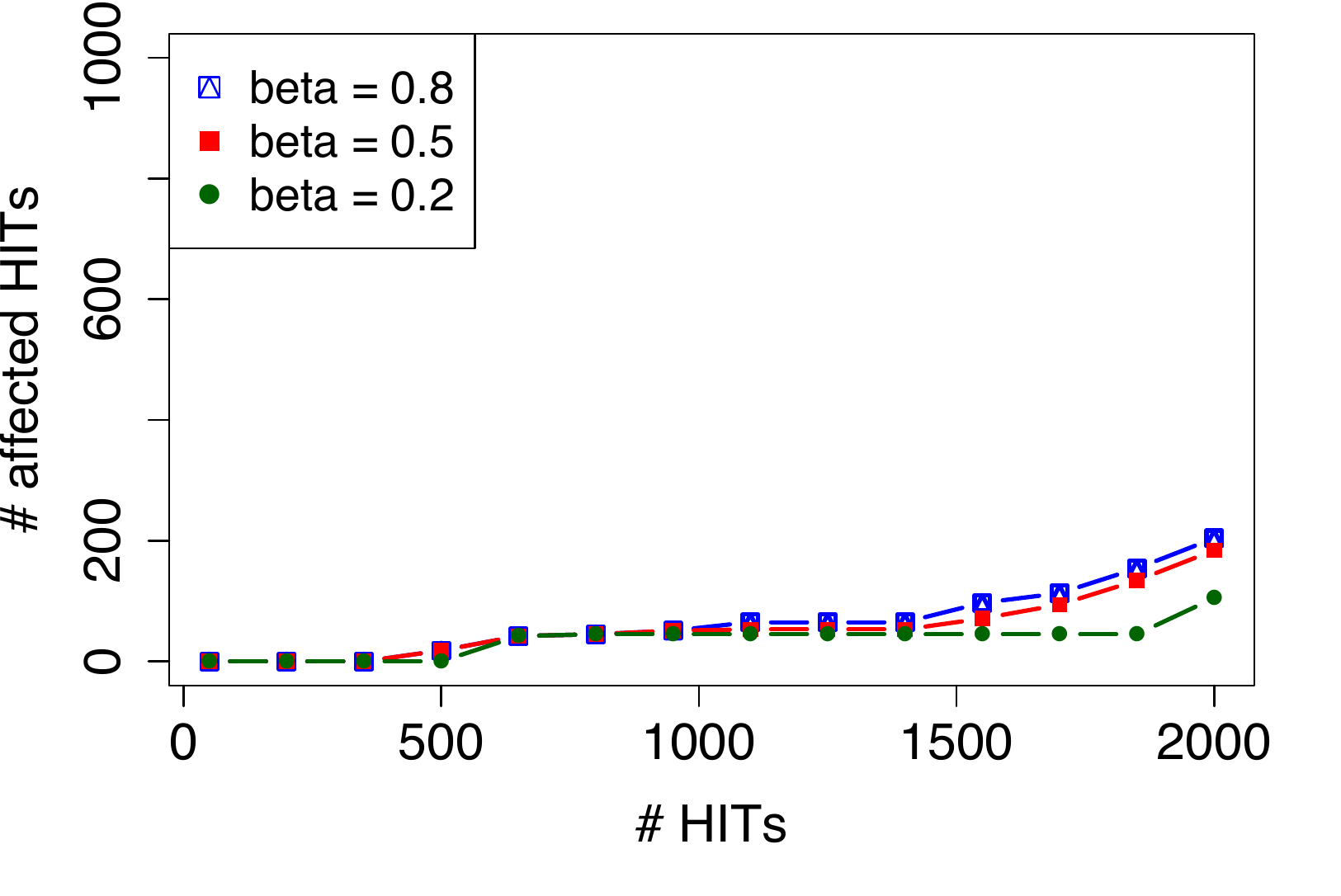} \\
    \begin{minipage}{.24\textwidth}
      \centering
      \vspace*{-5pt}
      {(a) States 2 experiment}
    \end{minipage} &
    \begin{minipage}{.24\textwidth}
      \centering
      \vspace*{-5pt}
      {(b) UN 2 experiment}
    \end{minipage} &
    \begin{minipage}{.24\textwidth}
      \centering
      \vspace*{-5pt}
      {(c) Ice Cream Flavors}
    \end{minipage}
  \end{tabularx}
 \vspace*{-10pt}
\caption{HITs detected as list-walking for different experiments}
\label{fig:listuse}
 \vspace*{-5pt}
\end{figure*}

The probability in equation~\ref{eqn:listsum} determines if the target sequence shared amongst $w$ out of $W$ workers is likely caused by list walking.
We now discuss $p_\alpha$, the probability of observing a particular target sequence $\alpha$ of length $s$.

\subsubsection{Defining the probability of a target sequence}
Not all workers use the same list or use the same order to walk through the list, so we want $p_\alpha$ to reflect the observed answer sequences from workers.  
We do this by estimating the probability $p_\alpha(i)$ of encountering answer $\alpha_i$ in the $i^{th}$ position of the target sequence by the fraction of times this answer appears in the $i^{th}$ position among all $W$ answers.
Let $\mf(i)$ be the number of times answer $\alpha_i$ appears in the $i^{th}$ position amongst all the sequences $W$ being compared, $p_\alpha(i)$ is defined as $\mf_i/W$. For example, if the target sequence $\alpha$ starting at offset $o$ is ``A,B,C'' and the first answers for four workers are ``A'',``A'',``A'', and ``B'', respectively, $\mf_{o + 1}/W$ would be $3/4$. 
Now the probability of seeing $\alpha$ is a product of the probabilities of observing $\alpha_{o+1}$, then $\alpha_{o+2}$, etc.
\begin{equation}
	p_\alpha = \prod_{i=o}^{o+s} \frac{\mf_i}{W}
\label{eqn:listsum2}
\end{equation}

Relying solely on the data in this manner could lead to false negatives in the extreme case where $w=W$, i.e., where all workers use the same target sequence. Note that in this case $p_\alpha$ attains the maximum possible value of $1$. As a result, $p_\alpha$ will be greater than any threshold we pick, and hence this case will be rejected as a chance occurrence.
What we really want is to incorporate \emph{both} the true data via $\mf_i/W$ as well as our most pessimistic belief of the underlying skew. 
As a pessimistic prior, we choose the highly skewed Gray’s self-similar distribution \cite{QuicklyGenerating}, often used for situations following the 80/20 rule.
That is, only if we find a sequence which can not be explained (e.g., with more than 1\% chance) with the 80/20 self similar distribution, we believe we have encountered list walking.
Assuming a high skew distribution is conservative because it is more likely that workers will answer in the same order if they were truly sampling than with, say, a uniform distribution.
The self-similar distribution with $h=0.2$ in particular is advantageous for our analysis because in the sampling without replacement paradigm, the most likely item has 80\% ($1-h=0.8$) chance of being selected and, once that item is selected and removed, the next most likely item has an 80\% chance as well.

As a first step, we assume that the target sequence follows the self-similar distribution exactly by always choosing the most likely sequence. In this case $\alpha$ is simply a concatenation of the most likely answer, followed by the second most likely answer, and so on.
Hence the likelihood of selecting this sequence under our prior belief is $(1-h)^{s}$ and the likelihood that a set of $w$ workers select this same sequence:
\begin{equation}
(1-h)^{sw} 
\label{eqn:mostlikily}
\end{equation}
Note that this probability does not calculate the probability of having \textit{any given} sequence of length $s$ shared among $w$ workers; instead it represents the likelihood of having the most likely sequence in common. 
Incorporating the probability of all sequences of length $s$ would be the sum of the probabilities of each sequence order, i.e., the most likely sequence $+$ the second most likely sequence, etc.
However, we found that the terms after the most likely sequence contribute little and our implementation of that version had little effect on the results; thus do not consider it further. 
%
%

To combine the distribution derived from data and our prior belief in the maximum skew, we introduce the smoothing factor $\beta$ to shift the emphasis from the data to the distribution; higher values of $\beta$ indicate putting more emphasis on the data.
Using $\beta$ to combine equation~\ref{eqn:listsum2} with equation~\ref{eqn:mostlikily}, we yield the probability of having the target sequence $\alpha$ (of length $s$) in common:
\vspace{-0.1in}
\begin{equation}
	p_\alpha = \prod_{i=1}^s \left(\beta \frac{\mf_i}{W} + (1 - \beta) (1-h)\right)
\label{eqn:p}
\end{equation}
If $\beta = 1$, $p_\alpha$ only incorporates the frequency information from the data, so if all workers are walking down the same list, then the probability in equation~\ref{eqn:p} would be $1$ (thus not detecting the list use).
Note also that when $\beta = 0$, $p_\alpha$ just uses the $80-20$ distribution and will reduce to $(1-h)^{s}$.
We demonstrate the effect of different values of $\beta$ next. 


\subsection{Experimental Results}
\label{sec:listwalking:result}
To apply our heuristic to the experiments run on AMT, we investigate sliding windows of length $s$, with $s \geq 5$ and up to the maximum sequence length from any worker.
For a given window of size $s$ that has more than one worker with the same sequence, we compute the probability of that sequence using equation~\ref{eqn:listprob} as described above.
If the probability falls below the threshold $0.01$, we consider the sequence as being from a list.
Our version of windowing ensures that we compare sequences that start at the same offset $o$ across all workers.
This makes sense for equation~\ref{eqn:p} which leverage the relative order that workers provide answers.
A shingling approach would provide more windows to compare across workers, and could thus detect list candidates at different offsets across workers, but our equations do not apply in this scenario.
Furthermore, the idea of checking for sequences that are exactly the same will suffer if a worker has a gap in part of his sequence.
However, we show below that our technique is effective in detecting list use.

For a given experiment, we are interested in both when list use can be detected as well how widespread it is.
We check for list use over time (number of HITs) and quantify how many of the observed HITs were part of a list; this gives a sense of the impact of list use in the experiment.
Due to limited space, we describe only a few of the experiments.

Figure~\ref{fig:listuse} shows the number of affected HITs in one of the States experiments, one of the UN experiments, and for the ice cream flavors experiment.
We use representative single runs opposed to averages to better visualize the effect what a user of the systems would observe. 
The lines correspond to using the equation~\ref{eqn:p} for the different $\beta$ values $0.2,0.5,0.8$.
In general, lower values of $\beta$ detect fewer lists or it takes more HITs to discover the lists.

The states experiments experienced little or no list walking. 
While there are definitely webpages that show the list of US states, perhaps it was not too much harder for workers to think of them on their own. 
All UN experiments exhibited some list use, with the list of course being the alphabetical list of countries that can be found online. 
Interestingly, we also detect some list walking in the ice cream experiment, despite it being a personal question easily answerable without consulting a source online. 
After some searching for the original sources, we actually found a few lists used for ice cream flavors, like those from the ``Penn State Creamery" and ``Frederick's Ice Cream".
Several lists were actually not alphabetical, including a list of the ``15 most popular ice cream flavors'' as well as forum thread on ChaCha.com discussing ice cream flavors. 

Our results show that our heuristic is able to detect when multiple workers are consulting the same list.
Furthermore, it is able to report in most cases on the impact of list walking on the overall result. 
For example, it reports that for the UN 2 experiment around 20-25\% of all HITs are impacted by list walking. 
Whereas for the ice cream flavors experiment less than 10\% are impacted. 
In both cases, the impact on the estimator was not significant.
However, in another UN experiment run we observed list walking that at times exceeded 40\%, and indeed in this experiment run the estimator under-predicted more than in the others (after 600 HITs it was still under-predicting the cardinality by 40). 
As future work, we plan to automatically correct the estimation with the knowledge of list walking as well as explore alternative crowdsourcing strategies.
 




\section{Related Work}
\label{sec:related}

In this paper we focused on estimating progress towards completion of a query result set, which is an aspect of query quality. 
To our knowledge, quality of an open-ended question posed to the crowd has not been directly addressed in crowdsourcing literature.
In contrast, various techniques have been proposed for quality control for individual set elements \cite{panos_quality, crowdtutorial}.

Our estimation techniques build on top of existing work on species or class estimation \cite{bunge_review93,colwell_review94,chao_review05}. 
These techniques have also been used in database literature for distinct value estimation as discussed in Section~\ref{sec:bg_est}. 

The database community has developed a recent interest in designing new database systems that incorporate crowdsourced information. 
The presented techniques here are not restricted to CrowdDB \cite{crowddb} and apply likewise to other hybrid human-machine database systems, such as Qurk or sCOOP. 
Qurk \cite{qurk} encapsulates crowd input using UDFs; task templates generate AMT HIT UIs for performing crowd tasks like verification, joins, and sorting, and specifying quality control algorithms like majority vote. 
Deco \cite{deco} (part of the sCOOP project \cite{scoop}) extends the internal schema with functional dependencies, as well as``fetch'' and ``resolution'' rules for crowdsourcing tuples and resolving conflicts, respectively. 
Both systems allow to require sets from the crowd and do not yet provide any quality control mechanisms for it. 

Finally, there exists a variety of literature on crowdsourcing in general, addressing issues from techniques to improve and control latency \cite{busywait,automan} to correcting the impact of different worker capabilities \cite{duckwork}. 
This work is orthogonal to estimating the quality of sets and not further discussed. 




\section{Future work and Conclusion}
\label{sec:future}
People are particularly well-suited for gathering new information because they have access to both real-life experience and online sources of information. 
Incorporating crowd-sourced information into a database, however, raises the question of what query results mean without the closed-world assumption -- how does one even reason about a simple \texttt{SELECT *} query?
In this paper, we showed how algorithms for species estimation can be applied to crowdsourced query results to  evaluate trade-offs between cost and completeness.
Although the standard estimators work surprisingly well, crowd-specific behavior can influences the quality of the completeness estimation.
We therefore developed two new heuristics: the first one corrects the sample for the effect of streakers, whereas the second heuristic detects list-walking and informs the user about the opportunity of changing the crowd-sourcing strategy.

Many future directions exist, ranging from different user interfaces for soliciting worker input to incorporating the above techniques into a query optimizer. 
We have done initial explorations into a ``negative suggest'' UI that only allows workers to enter new answers: workers are presented with the list of existing answers, and they cannot submit an answer that appears on that list. 
A hybrid approach using this interface coupled with our current interface could be used to grow a set and/or help find rare items. 
In this paper, we assumed that workers do not provide incorrect answers. 
The literature already proposed a variety of quality control solutions for single answers. 
However, fuzzy set membership (e.g., is Pizza or Basil a valid ice-cream flavor\footnote{Basil is actually quite a delicious ice cream flavor, but we doubt that Pizza is.}) imposes interesting new challenges on the quality control for sets. 
Finally, we plan to build a budget-based query optimizer for hybrid human-machine systems.

By using statistical techniques we enable users to reason about the query progress and decide on cost-benefit trade-offs even in the presence of the open-world.



\section{Acknowledgments}
This work was inspired by the CrowdDB project, and we would like to thank our CrowdDB collaborators Donald Kossmann, Sukriti Ramesh, and Reynold Xin. 

This research is supported in part by a National Science Foundation graduate fellowship, and by gifts from Google, SAP, Amazon Web Services, Blue Goji, Cloudera, Ericsson, General Electric, Hewlett Packard, Huawei, IBM, Intel, MarkLogic, Microsoft, NEC Labs, NetApp, Oracle, Quanta, Splunk, VMware and by DARPA (contract \#FA8650-11-C-7136).

\begin{small}
\bibliographystyle{abbrv}
\bibliography{vldb_sample}  
\end{small}


\balancecolumns

\end{document}